\documentclass[journal,draftcls,onecolumn,12pt,twoside]{IEEEtranTCOM}

\usepackage{comment}
\usepackage{cite}
\usepackage{graphicx}
\usepackage{epstopdf}
\usepackage{amsthm}
\usepackage[cmex10]{amsmath}
\usepackage{array}
\usepackage{float}
\usepackage{graphicx}
\usepackage{amssymb}
\usepackage{color}
\usepackage[font=footnotesize]{caption}
\usepackage{multirow}
\usepackage{balance}
\usepackage{amsmath}
\usepackage{hhline}

\newtheorem{theorem}{Theorem}
\newtheorem{lemma}{Lemma}
\newtheorem{definition}{Definition}
\newtheorem{remark}{Remark}
\newtheorem{example}{Example}

\bstctlcite{IEEEexample:BSTcontrol}

\newcommand\blfootnote[1]{
  \begingroup
  \renewcommand\thefootnote{}\footnote{#1}
  \addtocounter{footnote}{-1}
  \endgroup
}

\normalsize

\begin{document}

\title{Finite-Length Construction of High Performance Spatially-Coupled Codes via Optimized Partitioning and Lifting}

\author{Homa~Esfahanizadeh,~\IEEEmembership{Student Member,~IEEE,}
        Ahmed~Hareedy,~\IEEEmembership{Student Member,~IEEE,}
        and~Lara~Dolecek,~\IEEEmembership{Senior~Member,~IEEE}\vspace{-1.5cm}}

\markboth{IEEE Transactions on Communications}{Submitted paper}

\maketitle
\begin{abstract}
Spatially-coupled (SC) codes are a family of graph-based codes that have attracted significant attention thanks to their capacity approaching performance and low decoding latency. An SC code is constructed by partitioning an underlying block code into a number of components, and coupling their copies together.
In this paper, we first introduce a general approach for the enumeration of detrimental combinatorial objects in the graph of finite-length SC codes. Our approach is general in the sense that it effectively works for SC codes with various column weights and memories. Next, we present a two-stage framework for the construction of high performance binary SC codes optimized for additive white Gaussian noise channel; we aim at minimizing the number of detrimental combinatorial objects in the error floor regime. In the first stage, we deploy a novel partitioning scheme, called the optimal overlap partitioning, to produce optimal partitioning corresponding to the smallest number of detrimental objects. In the second stage, we apply a new circulant power optimizer to further reduce the number of detrimental objects in the lifted graph. 
An SC code constructed by our new framework has nearly 5 orders of magnitudes error floor performance improvement compared to the uncoupled setting.
\end{abstract}\vspace{-0.8cm}

\blfootnote{
\thanks{H. Esfahanizadeh, A. Hareedy, and L. Dolecek are with the Department of Electrical and Computer Engineering, University of California, Los Angeles, Los Angeles, CA 90095 USA (e-mail: \{hesfahanizadeh, ahareedy\}@ucla.edu; dolecek@ee.ucla.edu).}\\Parts of the paper were presented at the IEEE International Symposium on Information Theory (ISIT) 2017 \cite{ourISIT2017}, and the IEEE Information Theory Workshop (ITW) 2017 \cite{ourITW2017}. Research supported in part by a grant from ASTC-IDEMA and NSF-CCF grant No. CCF-1150212.}

\IEEEpeerreviewmaketitle

\vspace{-0.3cm}
\section{\vspace{-0.2cm}Introduction}
\vspace{-0.2cm}
\IEEEPARstart{S}{patially}-coupled (SC) codes are graph-based codes constructed by coupling together a series of disjoint block codes into a single coupled chain \cite{FelstromIT1999,IyengarIT2013}. This operation can be viewed as partitioning the parity-check matrix $\mathbf{H}$ of a block code into component matrices $\mathbf{H}_i$, where $i\in\{0,1,\cdots,m\}$, and piecing $L$ copies of component matrices together to obtain the parity-check matrix $\mathbf{H}_{SC}$ of an SC code, as shown in Fig.~1. The parameters $m$ and $L$ are called the memory and coupling length, respectively. Here, as the underlying block codes, we consider circulant-based (CB) LDPC codes, where all circulants are non-zero {\cite{TannerIT2004}}.

\begin{figure}\label{fig_SC_structure}
\vspace{-0.4cm}
\centering
\includegraphics[width=0.27\textwidth]{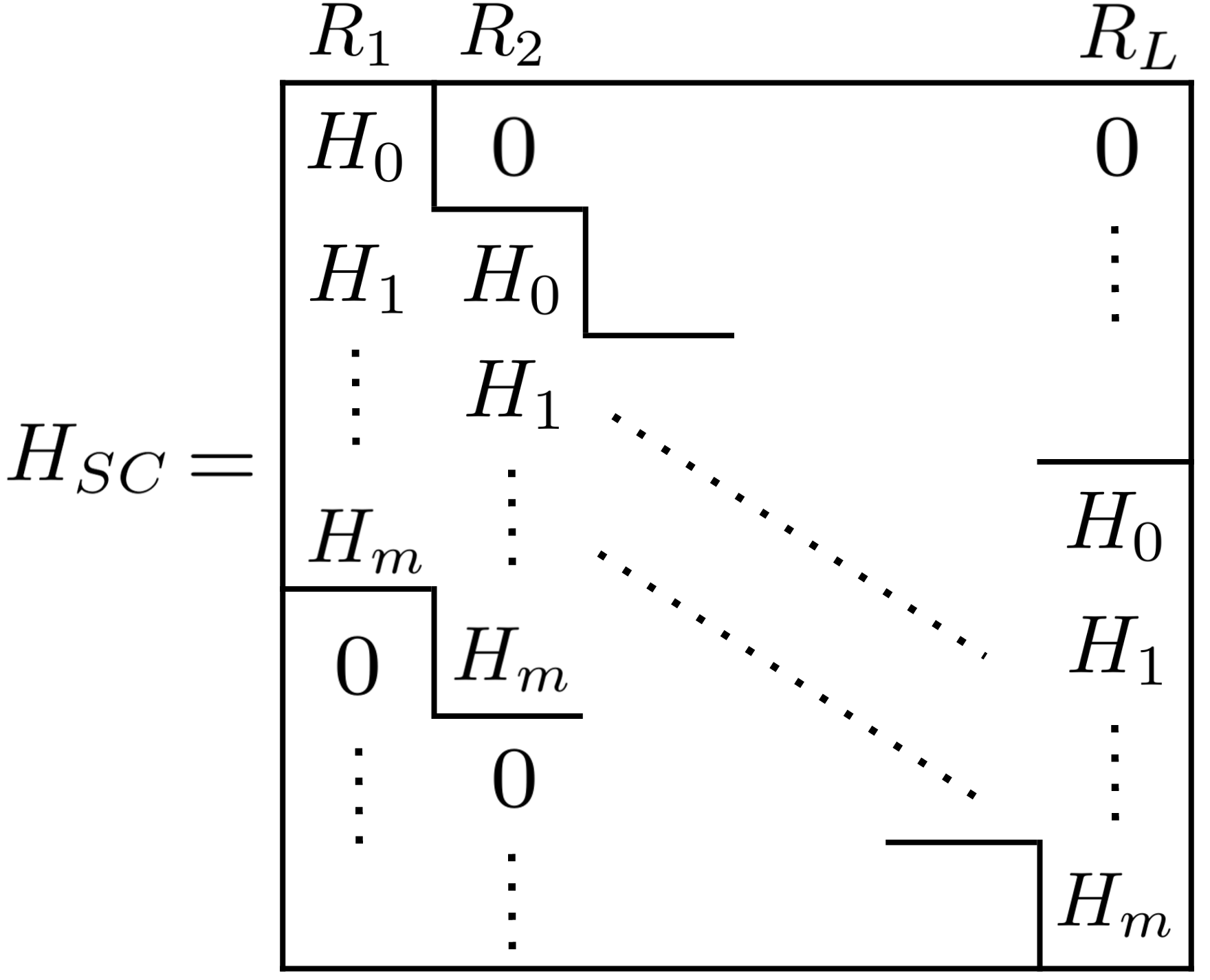}\vspace{-0.2cm}\\
\caption{\vspace{-0.5cm}The parity-check matrix of an SC code with parameters $L$ and $m$.}
\vspace{-1cm}
\end{figure}

Significant recent research on SC codes has been devoted to the asymptotic analysis, i.e., where the code length approaches infinity, e.g., \cite{LentmaierIT2010,KudekarIT2011,KudekarIT2013,CostelloISIT2014,AndriyanovaIT2016}. While the asymptotic analysis is important, the results cannot be immediately translated to the finite-length case, because the underlying assumptions on the code structure, e.g., being cycle-free, are different. 
There are several works that have studied the finite-length analysis and design of SC codes in the error floor area, e.g., \cite{MitchellISIT2014,AmiriTCOM2016,EsfahanizadehTMAG2017,MitchellISIT2017,SmarandacheISIT2017}.
These works, while promising, have some limitations: 
In \cite{MitchellISIT2014}, \cite{AmiriTCOM2016}, and \cite{EsfahanizadehTMAG2017}, authors consider SC codes with memory $m=1$ and focus on a restricted partitioning technique of cutting vectors. In \cite{MitchellISIT2017}, a construction method is presented for the the class of array-based (AB) SC codes with column weight $\gamma=3$ and for different memories. In \cite{SmarandacheISIT2017}, an approach is presented for improving the girth properties of SC codes.

In this paper, we propose a new combinatorial framework for the finite-length analysis and design of circulant-based SC (CB-SC) codes. We aim at constructing SC codes with the minimum number of problematic objects in the error floor regime. These problematic objects are certain configurations in the graph of codes that depend on both the code specifications as well as the channel model \cite{Richardson2003,DolecekIT2010,UrbankeIT2002,HareedyJSAC2016}. We first introduce a new enumeration approach that exploits the structure of SC codes in order to efficiently enumerate combinatorial objects of interest. Our new approach is more general than our previous work \cite{EsfahanizadehTMAG2017}, since it can be applied to SC codes constructed by an arbitrary partitioning and with memories $m\geq1$.

Next, we present a systematic scheme for partitioning the underlying block code and constructing SC codes with a superior performance in the error floor regime over additive white Gaussian noise (AWGN) channels. We operate on the protograph of the SC code, and express the number of subgraphs we want to minimize in terms of the overlap parameters, which characterize the partitioning. Then, we solve this discrete optimization problem to determine the optimal overlap parameters. We call this new partitioning scheme the optimal overlap (OO) partitioning. 

OO partitioning scheme is in particular suitable for code optimization in the regime outside the reach of brute force methods, since it finds the optimal partitioning in a systematic way and does not require to search among a set of possible choices. 
We demonstrate that the new scheme shows much better performance compared to the existing approach of partitioning by cutting vectors \cite{MitchellISIT2014,AmiriTCOM2016}. More importantly, our partitioning scheme is presented for general memory $m$ and column weight $\gamma$. Given the optimal partitioning, we then apply a new heuristic program to optimize the circulant powers of the underlying block code to further reduce the number of problematic objects. We call this heuristic program the circulant power optimizer (CPO).

The rest of the paper is organized as follows. In Section~II, we introduce the preliminaries. In Section~III, we propose our general enumeration approach. In Section~IV, we propose our optimal overlap partitioning scheme and circulant power optimizer to construct SC codes. Our simulation results are given in Section V. Finally, the conclusion appears in Section~VI.

\vspace{-0.7cm}
\section{Preliminaries}
\vspace{-0.3cm}
In this section, we describe CB-SC codes. Then, we review the definition of problematic objects causing the error floor degradation for SC/block codes over AWGN channels.
\vspace{-1cm}
\subsection{Circulant-Based SC Codes}
\vspace{-0.1cm}
CB codes are a class of structured regular $(\gamma,\kappa)$ LDPC codes, where $\gamma$ is the column weight of the parity-check matrix (variable node degree in the graph), and $\kappa$ is the row weight (check node degree). CB codes can offer simple hardware implementation thanks to their structure {\cite{TannerIT2004}}. Suppose $p$ is the size of the constituent circulants. The parity-check matrix $\mathbf{H}$ of a CB code is constructed as follows: 
\vspace{-0.3cm}\begin{equation*}
\mathbf{H}=\left[
\begin{array}{cccccccc}
\sigma^{f_{0,0}} & \sigma^{f_{0,1}}  & \dots & \sigma^{f_{0,\kappa-1}} \\
\sigma^{f_{1,0}} & \sigma^{f_{1,1}}  & \dots & \sigma^{f_{1,\kappa-1}} \\
\vdots & \vdots & \dots & \vdots\\
\sigma^{f_{\gamma-1,0}} & \sigma^{f_{\gamma-1,1}}  & \dots & \sigma^{f_{\gamma-1,\kappa-1}}
\end{array}
\right].
\end{equation*}

Here, $\sigma$ denotes the $p \times p$ circulant matrix obtained by cyclically shifting the columns of an identity matrix one unit to the left. Each column of $\mathbf{H}$ corresponds to one variable node (VN) and each row corresponds to one check node (CN). In the parity-check matrix $\mathbf{H}$, let $i$, $0 \leq i \leq \gamma-1$, be the row group index and $j$, $0 \leq j \leq \kappa-1$, be the column group index. The circulant powers are the non-negative integer values denoted by $f_{i,j}$.
For example, the choice of $f_{i,j}=ij$ and $\kappa=p$ results in the class of AB codes \cite{FanTurbo2000}.

SC codes have parity-check matrices, $\mathbf{H}_{SC}$, with a band-diagonal structure. A CB-SC code is constructed by partitioning the $\kappa \gamma$ circulants in the parity-check matrix $\mathbf{H}$ of a block code into component matrices $\mathbf{H}_i$, $0\leq i\leq m$, where $m$ is referred to as memory. Each component matrix $\mathbf{H}_i$ has the same size as $\mathbf{H}$. A component matrix contains a subset of circulants in $\mathbf{H}$, and the rest of its elements are zero. Every circulant in $\mathbf{H}$ is assigned to exactly one of the component matrices, and $\sum_{i=0}^{m}\mathbf{H}_i=\mathbf{H}$. Given the component matrices $\mathbf{H}_i$ and the coupling length $L$, one can construct $\mathbf{H}_{SC}$ as shown in Fig.~1.

Approaches for constructing SC codes with underlying structured block codes were previously proposed in \cite{MitchellISIT2014}, \cite{AmiriTCOM2016}, and \cite{MitchellISIT2017}. The work in \cite{MitchellISIT2014} and \cite{AmiriTCOM2016} is based on a cutting vector scheme. In this scheme, the underlying block code is partitioned via a so-called cutting vector ${\boldsymbol{\zeta}}=[\zeta_0\text{ }\zeta_1\text{ }\dots\text{ }\zeta_{\gamma-1}]$ into component matrices $\mathbf{H}_0$ and $\mathbf{H}_1$. The cutting vector ${\boldsymbol{\zeta}}$ is a vector of ascending natural numbers. {Matrix $\mathbf{H}_0$ is constructed by copying all circulants of $\mathbf{H}$ with row and column group indices in $\{(i,j)|j<\zeta_i\}$ to the same coordinates in $\mathbf{H}_0$, and setting all remaining elements of $\mathbf{H}_0$ to $0$. Matrix $\mathbf{H}_1$ is then simply $\mathbf{H}_1{=}\mathbf{H}{-}\mathbf{H}_0$.} The cutting vector partitioning approach could be generalized to construct SC codes with higher memories by using several cutting vectors to partition the underlying block code. In this paper, we introduce a new scheme for partitioning the block code that notably outperforms the cutting vector based method. A new partitioning scheme is recently introduced in \cite{MitchellISIT2017} which is developed for memories $m=1,2$; however, the contribution is limited to column weight $\gamma=3$ and AB codes as the underlying block codes. 
\vspace{-0.4cm}
\subsection{Combinatorial Objects of Interest}
\vspace{-0.1cm}
Under iterative decoding algorithms, certain structures in the graph of graph-based codes cause the error floor phenomenon. These structures are error-prone, and the errors resulting from them are not necessarily codeword errors. We review the key definitions of these objects.

\vspace{-0.2cm}
\begin{definition}
Consider a subgraph induced by a subset $\mathcal{V}$ of VNs in the graph of a binary LDPC code. The set $\mathcal{V}$ is said to be an $(a, b)$ trapping set (TS) if the size of $\mathcal{V}$ is $a$ and the number of odd degree CNs connected to $\mathcal{V}$ is $b$ \cite{Richardson2003,BanihashemiIT2014}.
\end{definition}

\vspace{-0.6cm}
\begin{definition}
Consider a subgraph induced by a subset $\mathcal{V}$ of VNs in the graph of a binary LDPC code. Let $\mathcal{C}^e$ be the set of even degree CNs connected to $\mathcal{V}$, and let $\mathcal{C}^o$ be the set of odd degree CNs connected to $\mathcal{V}$. The set $\mathcal{V}$ is said to be an $(a, b)$ absorbing set (AS) if the size of $\mathcal{V}$ is $a$, the size of $\mathcal{C}^o$ is $b$, and each VN in $\mathcal{V}$ is connected to strictly more CNs in $\mathcal{C}^e$ than in $\mathcal{C}^o$ \cite{DolecekIT2010}.
\end{definition}
\vspace{-0.2cm}
Fig.~2 depicts the configuration of $(3,3)$, $(4,2)$, and $(5,3)$ ASs which are problematic objects for SC/block codes with column weight $\gamma=3$ over AWGN channels. Clearly, the class of TSs subsumes the class of ASs. While TSs that are not ASs are usually harmless (as for the AWGN channel, these configurations are typically unstable under iterative decoding), we purposely recall the definition here. The reason is, as we shall see later in the paper, a systematic elimination of multiple problematic ASs can be deftly achieved by focusing on the elimination of the common sub-structure that these ASs share. In some cases, these common structures will be certain TSs that on their own do not appear as decoding errors.
\vspace{-0.2cm}

\vspace{-0.4cm}
\begin{figure}[H]
\centering
\begin{tabular}{ccc}
\includegraphics[width=0.089\textwidth]{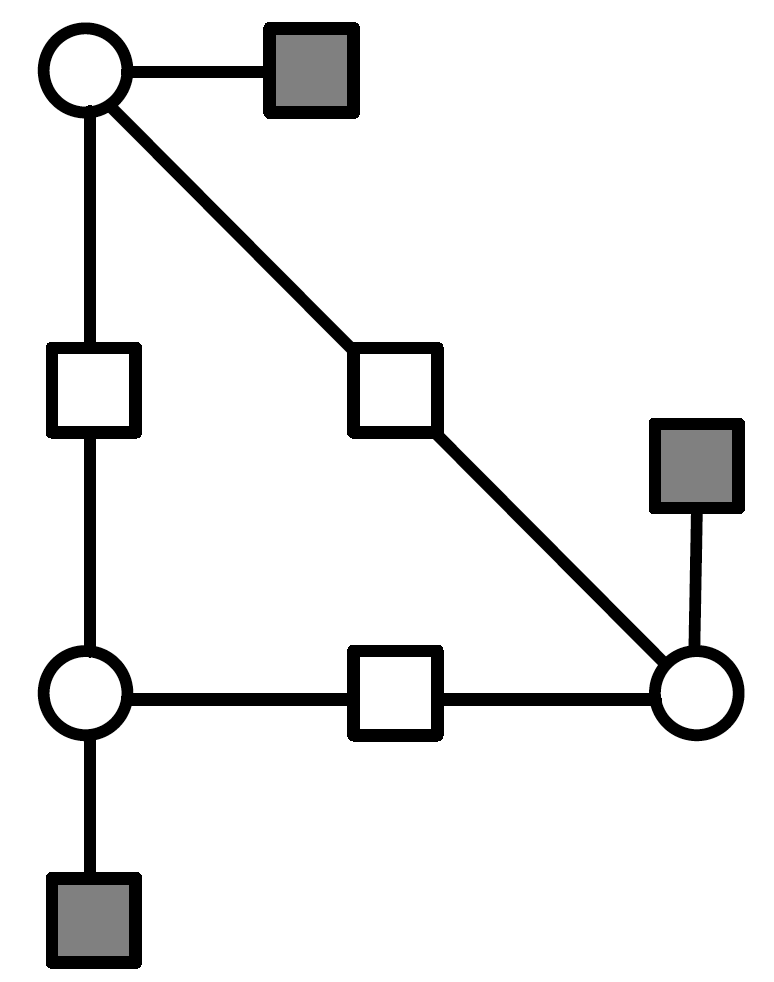}&
\includegraphics[width=0.089\textwidth]{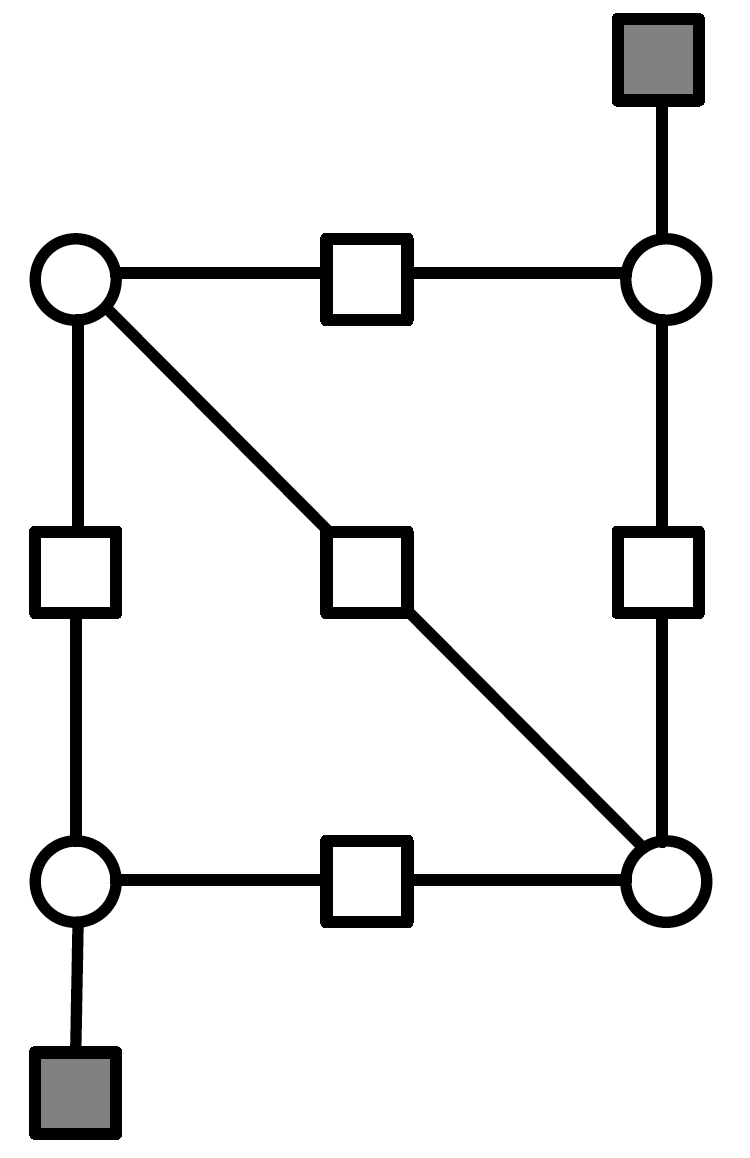}&
\includegraphics[width=0.191\textwidth]{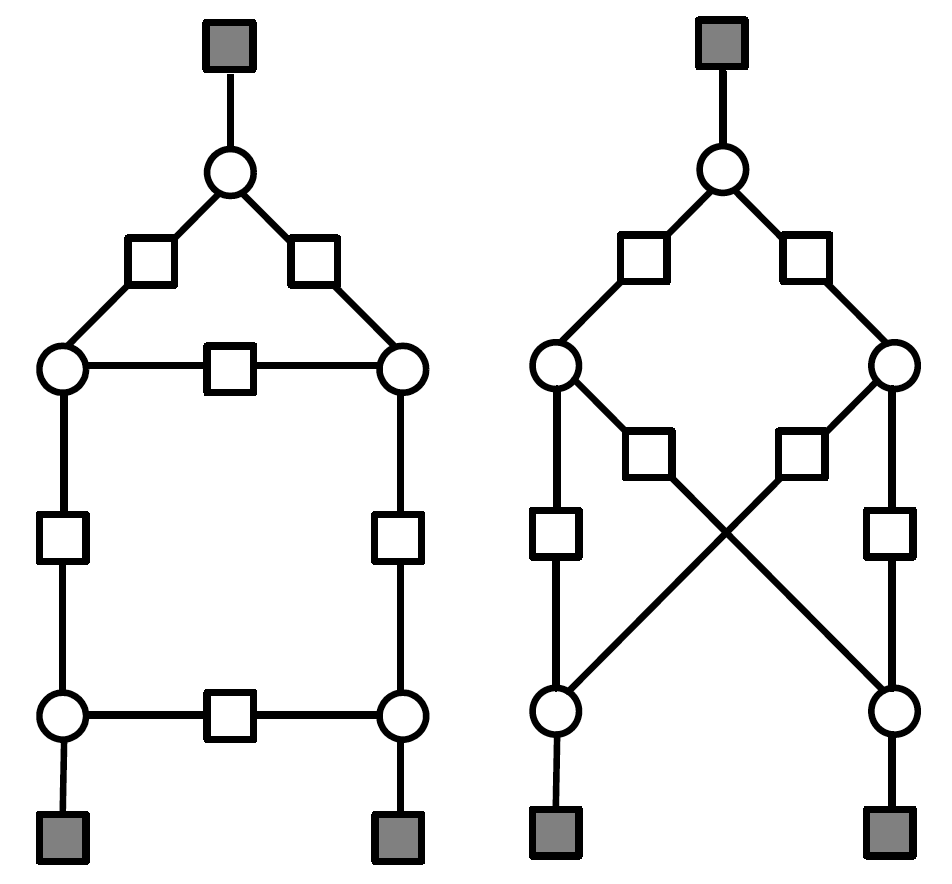}\vspace{-0.15cm}\\
(a)&(b)&(c)
\end{tabular}
\vspace{-0.2cm}
\caption{(a) $(3,3)$ AS; (b) $(4,2)$ AS; (c) two non-isomorphic configurations for $(5,3)$ AS.}
\vspace{-1cm}
\end{figure}
\vspace{-0.2cm}
\section{\vspace{-0.2cm}A General Approach for the Enumeration of Problematic Objects}

In this section, we introduce our new approach to enumerate combinatorial objects in the graph of SC codes. This approach can be applied to SC codes with any underlying CB code, partitioning, memory, and column weight.
Our main result is stated in Theorem~1. We first state the necessary auxiliary results and definitions.

\begin{definition}\label{def_replica}
\vspace{-0.2cm}
Consider an SC code with parameters $p$, $\kappa$, $\gamma$, $m$, and $L$. 
The replica $\mathbf{R}_r$, $r\in\{1,\cdots,L\}$, is a collection of columns in the matrix $\mathbf{H}_{SC}$ and is defined as:
\vspace{-0.2cm}\begin{equation}\label{equ_replica}
\vspace{-0.2cm}
	\mathbf{R}_r=\mathbf{H}_{SC}[0:(L+m)\gamma p-1][(r-1)\kappa p:r\kappa p-1].
\vspace{-0.2cm}
\end{equation}
\end{definition}

Fig.~1 illustrates the replicas on the parity-check matrix $\mathbf{H}_{SC}$ of an SC code. The notation $\mathbf{H}_{SC}[\rho_1:\rho_2][\nu_1:\nu_2]$ refers to a submatrix of $\mathbf{H}_{SC}$ with rows $\{\rho_1,\rho_1+1,\cdots,\rho_2\}$ and columns $\{\nu_1,\nu_1+1,\cdots,\nu_2\}$. Because of the structure of SC codes, the non-zero part of $\mathbf{R}_r$ is the same for any $r\in\{1,\cdots,L\}$:
\begin{equation}\label{equ_nz_replica}\vspace{-0.2cm}
\mathbf{R}_r[(r-1)\gamma p:(r+m)\gamma p-1][0:\kappa p-1]=[\mathbf{H}_0^T \mathbf{H}_1^T \cdots \mathbf{H}_{m}^T]^T.\vspace{-0.2cm}
\end{equation}

\begin{remark}\label{remark_graph_matrix_correspondance}
Each VN corresponds to one unique column and each CN corresponds to one unique row in the parity-check matrix. We thus interchangeably say that an $(a,b)$ AS/TS exists in the matrix/ exists in the graph of the code.
\vspace{-0.3cm}
\end{remark}

Let the shortest path that connects any two VNs of an $(a,b)$ AS/TS include at most $\lambda$ VNs. Given the configuration of an AS/TS, one can find the parameter $\lambda$ by known methods, e.g., Dijkstra's algorithm. In the case that there exists at least one cycle that spans all VNs, $\lambda$ is upper-bounded by $\left\lfloor \frac{a}{2} \right\rfloor+1$ (see \cite{EsfahanizadehTMAG2017}). We say two VNs are \textit{adjacent} if they are connected via a CN.
A $(3,3(\gamma-2))$ AS/TS has $\lambda=2$, because any two VNs are adjacent.

\begin{lemma}\label{lemma_chi}\vspace{-0.2cm}
For an SC code with memory $m$, all VNs of an $(a,b)$ AS/TS belong to at most $\chi$ consecutive replicas, where\vspace{-0.4cm}
\begin{equation}\label{equ_chi}\vspace{-0.2cm}
\chi=(\lambda-1)m + 1.
\vspace{-0.2cm}\end{equation}
\end{lemma}
\begin{IEEEproof}
As we can see in Fig.~1, the maximum number of consecutive replicas with the property that their non-zero parts have some row group indices in common is $(m+1)$. As a result, any two adjacent VNs must be within at most $(m+1)$ consecutive replicas. Among the VNs of the $(a,b)$ AS, let $v_1$ be the VN with the lowest index (the index of the corresponding column in $\mathbf{H}_{SC}$), and $v_f$ be the VN with the highest index. These two VNs are connected on a shortest path that includes at most $\lambda$ VNs. These VNs belong to at most $\lambda$ different replicas, and there can be at most $(m-1)$ different replicas between the replicas in which two adjacent VNs exist. Consequently, $v_1$ and $v_f$ must belong to a window of at most $\chi$ consecutive replicas, where $\chi$ is defined as follows: $\chi=(\lambda-1)(m-1)+\lambda=(\lambda-1)m + 1$. The rest of VNs of the AS/TS must also belong to these $\chi$ consecutive replicas since they have indices between $v_1$ and $v_f$. 
\end{IEEEproof}

We say an $(a,b)$ AS/TS \textit{starts} in replica $\mathbf{R}_r$ if among all it's VNs, the one with the lowest associated column index belongs to $\mathbf{R}_r$.\vspace{-0.2cm}
\begin{definition}\label{def_pi}\vspace{-0.2cm}
The matrix $\mathbf{\Pi}_r^k$, $r\in\{1,\cdots,L\}$ and $k\in\{1,\cdots,L-r+1\}$, is a submatrix of $\mathbf{H}_{SC}$, and is defined as follows:\vspace{-0.2cm}
\begin{equation}\label{equ_pi}
\begin{array}{lll}
&\mathbf{\Pi}_r^k=\mathbf{H}_{SC}[u_{r,1}^k:u_{r,2}^k][v_{r,1}^k:v_{r,2}^k],&\\
&u_{r,1}^k=(r-1)\gamma p,&u_{r,2}^k=(r+m+k-1)\gamma p-1,\\
&v_{r,1}^k=(r-1)\kappa p,&v_{r,2}^k=(r+k-1)\kappa p-1.
\end{array}
\end{equation}
\end{definition}
\begin{lemma}\label{lemma_AS_pi}\vspace{-0.2cm}
For an $(a,b)$ AS/TS that starts in replica $\mathbf{R}_r$ and spans $k$ consecutive replicas, all VNs and all their neighboring CNs have corresponding row and column indices within the submatrix $\mathbf{\Pi}_r^k$ of $\mathbf{H}_{SC}$.\vspace{-0.48cm}
\end{lemma}
\begin{IEEEproof}
An AS/TS that starts in replica $\mathbf{R}_r$ and spans $k$ consecutive replicas must have its VNs in the replicas $\{\mathbf{R}_r,\cdots,\mathbf{R}_{r+k-1}\}$. Based on (\ref{equ_replica}) and (\ref{equ_nz_replica}), the smallest submatrix of $\mathbf{H}_{SC}$ that spans all non-zero parts of these $r$ replicas is $\mathbf{\Pi}_r^k$.
\end{IEEEproof}
\vspace{-0.1cm}
\begin{theorem}\label{theorem_enumeration}\vspace{-0.2cm}
Consider an SC code with parameters $m$ and $L$. Let $F$ be the total number of $(a,b)$ ASs/TSs, and $F_1^k$ be the number of $(a,b)$ ASs/TSs that start in $\mathbf{R}_1$ and span $k$ consecutive replicas,\vspace{-0.3cm}
\begin{equation}\label{equ_enumeration}\vspace{-0.2cm}
F=\sum_{k=1}^{\chi}(L-k+1)F_1^k.\vspace{-0.2cm}
\end{equation}\vspace{-0.2cm}
\end{theorem}\vspace{-0.6cm}
\begin{IEEEproof}
By summing up the number of $(a,b)$ ASs/TSs over all possible starting replicas and spanning sizes, the total number of $(a,b)$ ASs/TSs can be written as:
\begin{equation}\label{equ_enu_splitting}
F=\sum_{k=1}^{\chi}\sum_{r=1}^{L-k+1}F_r^k.
\end{equation}
According to Lemma~\ref{lemma_AS_pi}, $F_r^k$ is equivalent to the number of objects of interest inside matrix $\mathbf{\Pi}_r^k$. Consider the matrix $\mathbf{\Pi}_{r+1}^k$, $r\in\{1,\cdots,L-k\}$:\vspace{-0.2cm}
\begin{align*}
&\mathbf{\Pi}_{r+1}^k=\mathbf{H}_{SC}[u_{r+1,1}^k:u_{r+1,2}^k][v_{r+1,1}^k:v_{r+1,2}^k],&\\
&u_{r+1,1}^k=u_{r,1}^k+\gamma p,&u_{r+1,2}^k=u_{r,2}^k+\gamma p,\\
&v_{r+1,1}^k=v_{r,1}^k+\kappa p,&v_{r+1,2}^k=v_{r,2}^k+\kappa p.\vspace{-0.2cm}
\end{align*}
Because of the repetitive structure of SC codes (see also Fig.~1), the following equality holds for the parity-check matrix $\mathbf{H}_{SC}$:\vspace{-0.2cm}
\begin{equation*}\vspace{-0.2cm}
\mathbf{H}_{SC}[i+p\gamma][j+p\kappa]=\mathbf{H}_{SC}[i][j].\vspace{-0.1cm}
\end{equation*}
Then,\vspace{-0.2cm}
\begin{equation*}\vspace{-0.2cm}
\mathbf{\Pi}_{r+1}^k=\mathbf{H}_{SC}[u_{r,1}^k+\gamma p:u_{r,2}^k+\gamma p][v_{r,1}^k+\kappa p: v_{r,2}^k+\kappa p]=\mathbf{H}_{SC}[u_{r,1}^k:u_{r,2}^k][v_{r,1}^k: v_{r,2}^k]=\mathbf{\Pi}_r^k.
\end{equation*}
Consequently,
\begin{equation*}\vspace{-0.3cm}
\mathbf{\Pi}_{r+1}^k=\mathbf{\Pi}_{r}^k \Longrightarrow F_{r+1}^k=F_r^k.\vspace{-0.2cm}
\end{equation*}
By means of induction, we can infer that:\vspace{-0.2cm}
\begin{equation}\label{equ_enu_induction}\vspace{-0.2cm}
F_r^k=F_1^k\hspace{1cm}\forall r\in\{1,2,\cdots,L-k+1\}.\vspace{-0.1cm}
\end{equation}
Combining (\ref{equ_enu_splitting}) and (\ref{equ_enu_induction}), yields the final result in (\ref{equ_enumeration}).
\end{IEEEproof}
The utility of Theorem~\ref{theorem_enumeration} is to significantly reduce the search size by seeking the enumeration over  $\mathbf{\Pi}_1^k$, $k\in\{1,\cdots,\chi\}$, rather than $\mathbf{H}_{SC}$; the latter can be easily computed even with an exhaustive search. In Section IV, we present a new scheme to efficiently find the quantities $F_1^k$'s for the protograph of SC codes (SC codes with $p=1$).
\vspace{-0.3cm}
\section{Finite-Length Construction of SC codes}
\vspace{-0.1cm}

For CB codes simulated over AWGN channels, certain types of ASs are dominant in the error floor regime.  For $\gamma=3$ codes, $(3,3)$, $(4,2)$, and $(5,3)$ ASs are the dominant objects \cite{AmiriTCOM2016,JiadongIT2013}, see Fig.~2. For $\gamma=4$ codes, the dominant objects are $(4,4)$ and $(6,4)$ ASs \cite{AmiriTCOM2016,JiadongIT2013}. There are several non-isomorphic configurations for the $(6,4)$ AS. The configuration of $(4,4)$ AS and one of the configurations of $(6,4)$ AS are illustrated in Fig.~3. For $\gamma=5$ codes, the dominant objects are $(4,8)$ and $(8,6)$ ASs \cite{JiadongIT2013}. The configuration of $(4,8)$ AS and one of the configurations of $(8,6)$ AS are illustrated in Fig.~4.

According to Figs~2-4, our extensive simulations, and the literature \cite{AmiriTCOM2016,JiadongIT2013}, {$(3,3(\gamma-2))$} AS/TSs appear as the subgraph of dominant problematic configurations for CB codes with column weight $\gamma$. We call the $(3,3(\gamma-2))$ AS/TS the \textit{common denominator}, which is a cycle of length~$6$ (cycle-$6$).
For $\gamma=3$, $4$, and $5$, the common denominators are $(3,3)$ AS, $(3,6)$ TS, and $(3,9)$ TS, respectively.
In order to design high performance SC codes, we seek to significantly reduce the number of dominant ASs. To efficiently perform this task, we aim at minimizing the number of common denominator instances. In fact, by minimizing the population of the common denominator instances as sub-structures, we reduce the number of all super-structures and improve the code performance. Moreover, the common denominator has a simpler combinatorial characteristics, and thus it is easier to locate and operate on.

\begin{figure}
\centering
\begin{tabular}{ccc}
\includegraphics[height=0.17\textwidth]{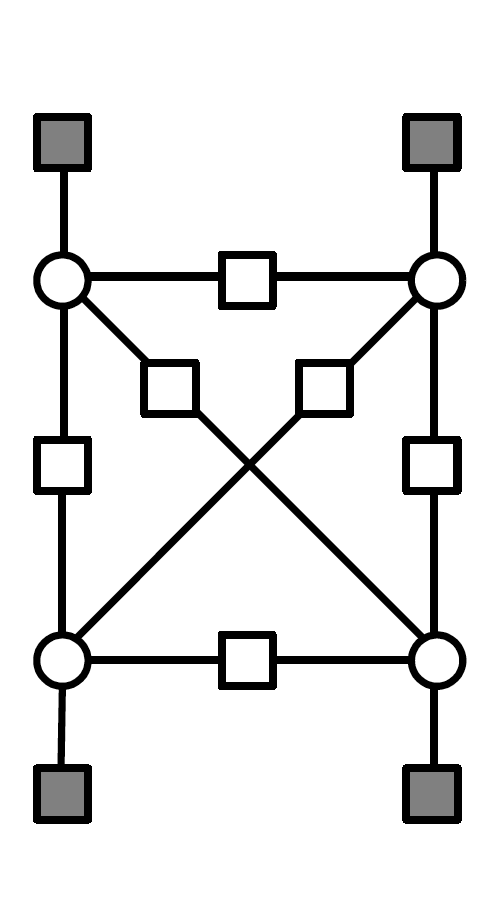}&
\includegraphics[height=0.17\textwidth]{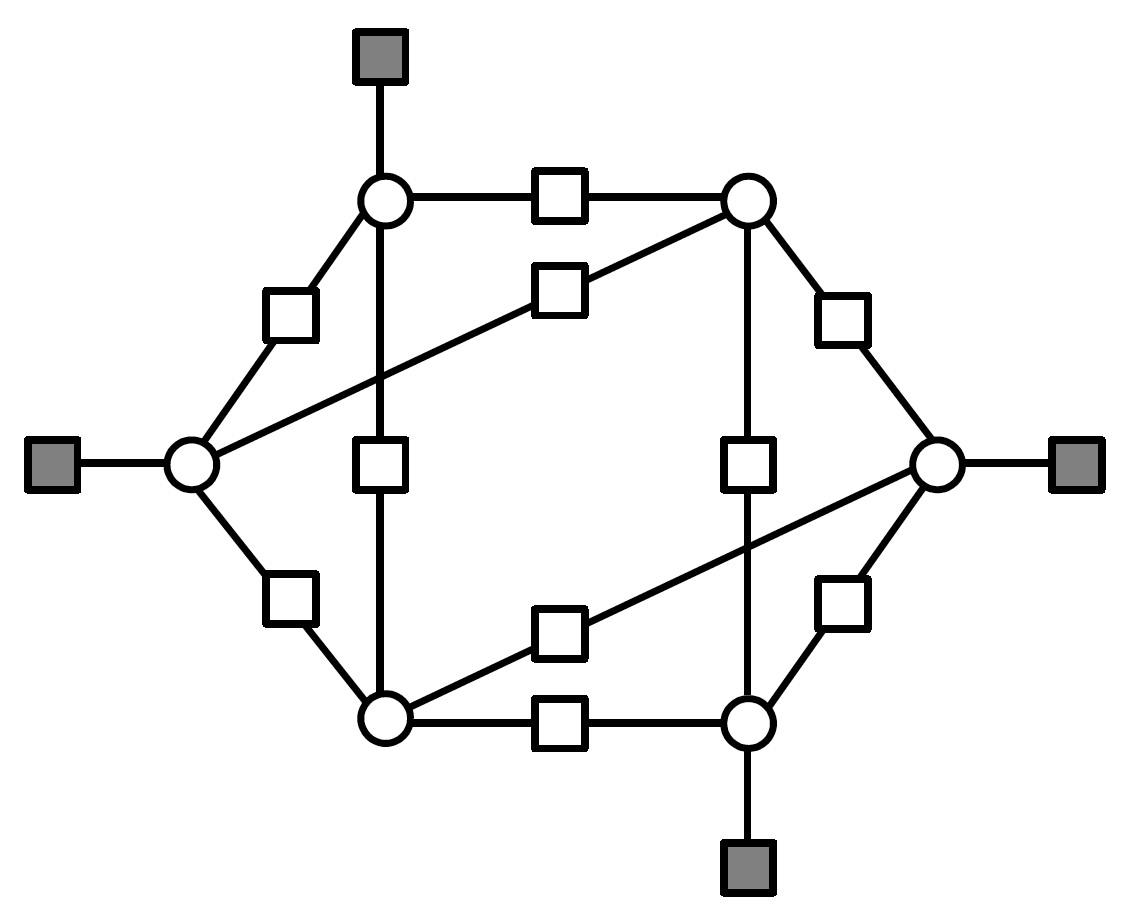}&
\includegraphics[height=0.17\textwidth]{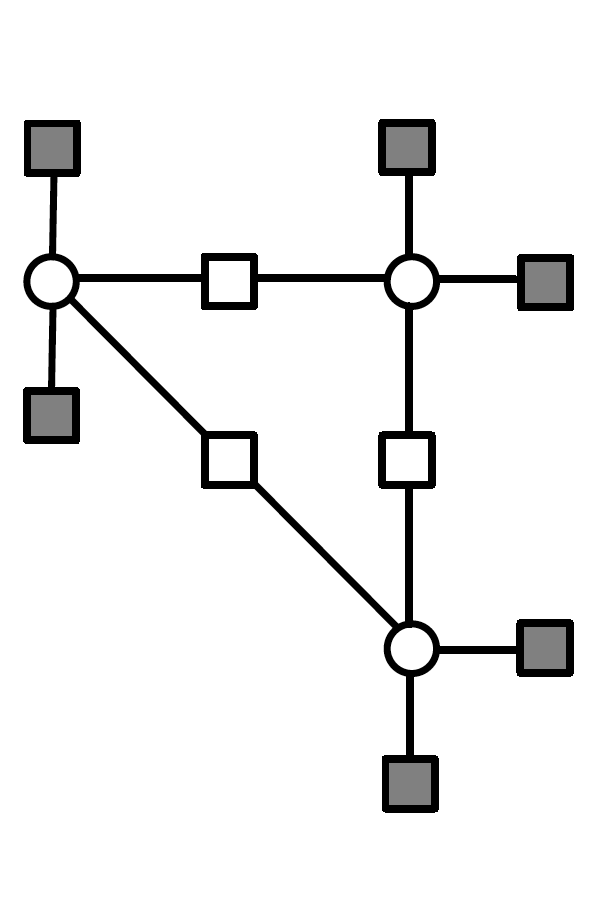}\vspace{-0.2cm}\\
(a)&(b)&(c)
\end{tabular}
\vspace{-0.3cm}
\caption{(a) $(4,4)$ AS; (b) one configuration for $(6,4)$ AS; (c) $(3,6)$ TS as the common denominator\vspace{-0.3cm}.}
\vspace{-0.2cm}
\end{figure}

\begin{figure}
\centering
\begin{tabular}{ccc}
\includegraphics[height=0.19\textwidth]{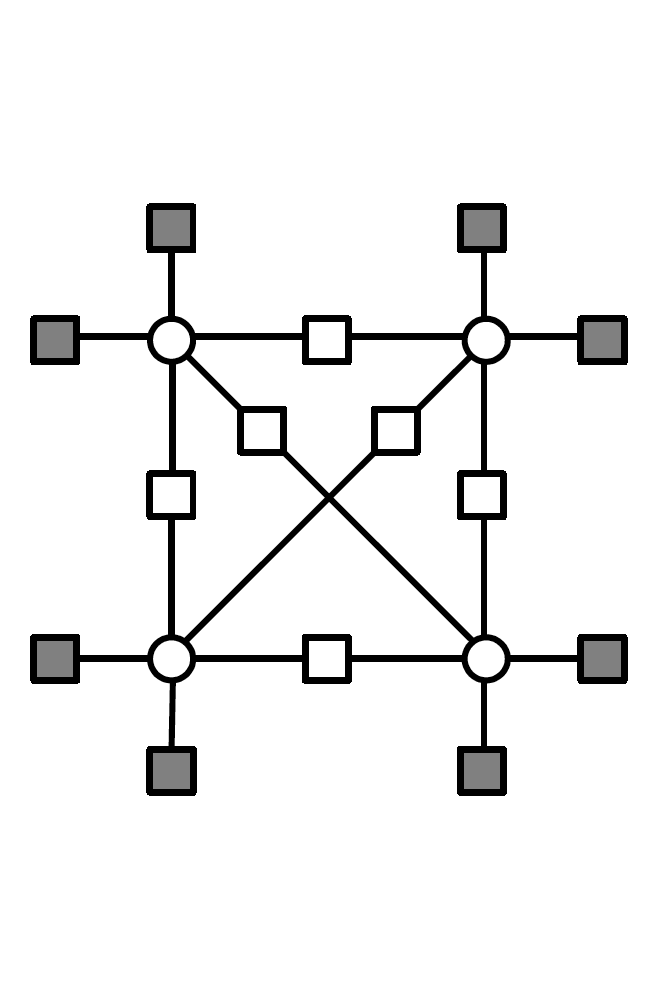}&
\includegraphics[height=0.19\textwidth]{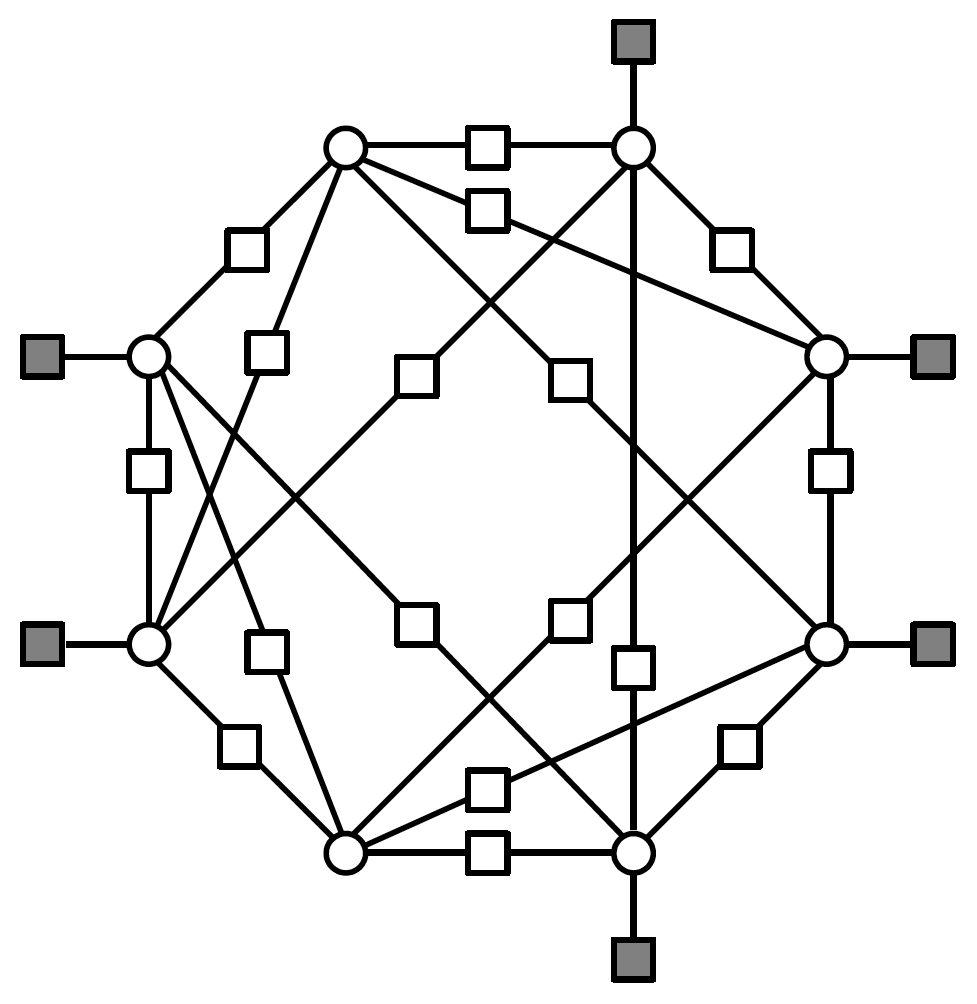}&
\includegraphics[height=0.19\textwidth]{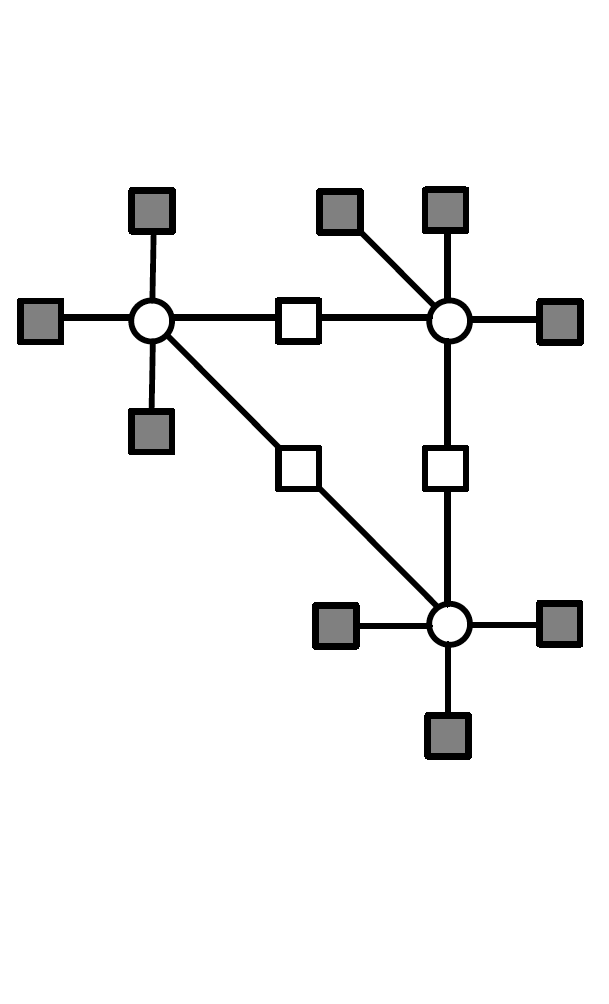}\vspace{-0.2cm}\\
(a)&(b)&(c)
\end{tabular}
\vspace{-0.3cm}
\caption{(a) $(4,8)$ AS; (b) one configuration for $(8,6)$ AS; (c) $(3,9)$ TS as the common denominator\vspace{-0.5cm}.}
\vspace{-0.4cm}
\end{figure}
In this section, we present a two-stage framework to design high performance SC codes.
\begin{enumerate}
\vspace{-0.2cm}
\item In the first stage, we operate on the protograph of the SC code, and express the number of cycles-$6$ in terms of the overlap parameters, which characterize the partitioning of the block code. Then, we solve this discrete optimization problem to determine the optimal overlap parameters. We call this new partitioning scheme the optimal overlap (OO) partitioning.
\item In the second stage, given the optimal partitioning obtained in the first stage, we apply a new heuristic program to optimize the circulant powers of the underlying block code to further reduce the number of cycles-$6$ in the graph of the SC code. We call this heuristic program the circulant power optimizer (CPO).
\end{enumerate}
The next two subsections describe these stages in details.
\vspace{-0.4cm}
\subsection{The Optimal Overlap Partitioning Scheme}
\vspace{-0.1cm}
We propose a novel combinatorial scheme, called \textit{Optimal Overlap} (OO), for partitioning the underlying block code and constructing SC codes. In OO partitioning, the resulting components do not necessarily each comprise a contiguous set of circulants. This property is in contrast with the scheme of partitioning with cutting vectors in which components -- by design -- have large overlaps \cite{MitchellISIT2014,AmiriTCOM2016}, which is an undesirable feature for the finite-length design as we show later. For constructing SC codes, we partition $\kappa\gamma$ circulants of an underlying code into $(m+1)$ components, and piece copies of them together to construct an SC code. The ultimate goal in the partitioning step is to construct SC codes that have the minimum number of problematic objects (cycles-$6$). This is achieved via a careful selection of the overlap parameters, as we show here.

The protograph of a CB matrix, $\mathbf{H}^p$, is the matrix resulting from replacing each $p\times p$ non-zero circulant in $\mathbf{H}$ with $1$, and each $p\times p$ zero circulant with $0$. The protographs of $\mathbf{H}_0$, $\mathbf{H}_1$, $\cdots$, $\mathbf{H}_m$ are $\mathbf{H}_0^p$, $\mathbf{H}_1^p$, $\cdots$, $\mathbf{H}_m^p$, respectively, and they are all of size $\gamma\times\kappa$. The protograph of $\mathbf{H}_{SC}$ is $\mathbf{H}_{SC}^p$, and it is of size $(L+m)\gamma\times L\kappa$. This $\mathbf{H}_{SC}^p$ also has $L$ replicas, $\mathbf{R}_r$, $1\leq r\leq L$, but with $1\times1$ circulants. The procedure of generating $\mathbf{H}_{SC}$ from $\mathbf{H}_{SC}^p$ is called lifting.

A cycle-$6$ in the graph of $\mathbf{H}_{SC}^p$ (the protograph of the SC code), which is defined by the non-zero
entries $\{(h_1,l_1),(h_1,l_2),(h_2,l_2),(h_2,l_3),(h_3,l_3),(h_3,l_1)\}$ in $\mathbf{H}_{SC}^p$, results in $p$ cycles-$6$ in the graph of $\mathbf{H}_{SC}$ if and only if \cite{Fossorier2004,6691250}:\vspace{-0.24cm}
\begin{equation}\label{equ_CP}
f_{h_{1},l_{1}}+f_{h_{2},l_{2}}+f_{h_{3},l_{3}}\equiv f_{h_{1},l_{2}}+f_{h_{2},l_{3}}+f_{h_{3},l_{1}} \text{ (mod $p$)},\vspace{-0.14cm}
\end{equation}
where $f_{h,l}$ is the power of the circulant indexed by $(h,l)$ in $\mathbf{H}_{SC}^p$ . Otherwise, this cycle results in $0$ cycle-$6$ in the graph of $\mathbf{H}_{SC}$ \cite{Fossorier2004,6691250}. Moreover, a cycle-$6$ in the final (lifted) graph of an SC code can only be generated from a cycle-$6$ in the protograph.

Motivated by the above fact, our OO partitioning aims at deriving the overlap parameters of $\mathbf{H}^p$ that result in the minimum number of common denominator instances in the graph of $\mathbf{H}_{SC}^p$, which is the protograph of the SC code. Then, we run the CPO to further reduce the number of common denominator instances in the graph of $\mathbf{H}_{SC}$ by breaking the condition in (\ref{equ_CP}) for as many cycles in the optimized graph of $\mathbf{H}_{SC}^p$ as possible. 

The goal here is to minimize the number of cycles-$6$ in the protograph of the SC code via the OO partitioning of $\mathbf{H}^p$, which is also the OO partitioning of $\mathbf{H}$. To achieve this goal, we establish a discrete optimization problem by expressing the number of cycles-$6$ in the graph of $\mathbf{H}_{SC}^p$ as a function of the overlap parameters and standard code parameters. We first introduce the overlap parameters:

\begin{definition}\label{def_ov_par}\vspace{-0.4cm}
From Definition~\ref{def_pi}, $\mathbf{\Pi}_1^{1,p}$ of size $(m+1)\gamma\times\kappa$ is the protograph of matrix  $\mathbf{\Pi}_1^1=[\mathbf{H}_0^T\cdots\mathbf{H}_m^T]^T$. A degree-$d$ overlap $t_{\{i_1,\cdots,i_d\}}$, $0\leq i_1,\cdots,i_d <(m+1)\gamma$, is defined as the overlap among $d$ distinct rows of \text{ }$\mathbf{\Pi}_1^{1,p}$ specified by the set $\{i_1,\cdots,i_d\}$, i.e., the number of positions in which all the $d$ rows have $1$'s simultaneously. A degree-$1$ overlap $t_{i_1}$, $0\leq i_1 < (m+1)\gamma$, is defined as the number of $1$'s in row $i_1$ of $\mathbf{\Pi}_1^{1,p}$.\vspace{-0.3cm}
\end{definition}
Remarks~\ref{remark_zero_ov_par_1} and \ref{remark_zero_ov_par_2} discuss the properties of the overlap parameters in Definition~\ref{def_ov_par}.\vspace{-0.1cm}
\begin{remark}\label{remark_zero_ov_par_1}\vspace{-0.2cm}
For an SC code with column weight $\gamma$, the maximum degree for an overlap parameter with non-zero value is $\gamma$. This is simply because there are exactly $\gamma$ $1$'s in any column of $\mathbf{\Pi}_1^{1,p}$, thus there is no position in a set of $d>\gamma$ rows such that all the rows have $1$'s simultaneously.\vspace{-0.2cm}
\end{remark}\vspace{-0.2cm}
\begin{remark}\label{remark_zero_ov_par_2}\vspace{-0.3cm}
Consider a set of rows  $\{i_1,\cdots,i_d\}$ of \text{ }$\mathbf{\Pi}_1^{1,p}$ and $d>1$. If there is at least one pair of distinct row indices $(i_u,i_v)$ such that $i_u,i_v\in\{i_1,\cdots,i_d\}$ and $i_u\equiv i_v \text{ (mod $\gamma$)}$; then, $t_{\{i_1,\cdots,i_d\}}=0$. This is because of the fact that the matrix $\mathbf{H}^p$ is partitioned into $\mathbf{H}_0^p$, $\cdots$, $\mathbf{H}_m^p$, thus there is zero overlap between similar rows of (protographs of) component matrices.\vspace{-0.3cm}
\end{remark}
Based on Definition~\ref{def_ov_par} and Remarks~\ref{remark_zero_ov_par_1} and \ref{remark_zero_ov_par_2}, the set of non-zero overlap parameters is:\vspace{-0.2cm}
\begin{equation}\label{equ_nz_ov_par}\vspace{-0.2cm}
\mathcal{O}=\{t_{\{i_1,\cdots,i_d\}}|1\leq d\leq \gamma,0\leq i_1,\cdots,i_d<(m+1)\gamma,\forall \{i_u,i_v\}\subset\{i_1,\cdots,i_d\}\hspace{0.1cm}i_u\not\equiv i_v \text{ (mod $\gamma$)}\}.\vspace{-0.2cm}
\end{equation}
\begin{example}\label{example_nz_ov_par}\vspace{-0.3cm}
For an SC code with $\gamma=3$ and $m=1$, the set of non-zero overlap parameters is:\vspace{-0.1cm}
\begin{equation*}
\begin{split}
\mathcal{O}&=\{t_{\{i_1,\cdots,i_d\}}|1\leq d\leq 3,0\leq i_1,\cdots,i_d<6,\forall \{i_u,i_v\}\subset\{i_1,\cdots,i_d\}\hspace{0.1cm}i_u\not\equiv i_v \text{ (mod $3$)}\}\\
&=\{t_0,t_1,t_2,t_3,t_4,t_5,t_{\{0,1\}},t_{\{0,2\}},t_{\{0,4\}},t_{\{0,5\}},t_{\{1,2\}},t_{\{1,3\}},t_{\{1,5\}},t_{\{2,3\}},t_{\{2,4\}},
t_{\{3,4\}},t_{\{3,5\}},t_{\{4,5\}},\\
&\hspace{0.8cm}t_{\{0,1,2\}},t_{\{0,1,5\}},t_{\{0,2,4\}},t_{\{0,4,5\}},t_{\{1,2,3\}},t_{\{1,3,5\}},t_{\{2,3,4\}},t_{\{3,4,5\}}\}.
\end{split}
\end{equation*}
There are $26$ non-zero overlap parameters.
\end{example}\vspace{-0.4cm}
The non-zero overlap parameters in (\ref{equ_nz_ov_par}) are not independent. In fact, some overlap parameters are linear combinations of other overlap parameters. Lemma~\ref{lemma_ind_nz_ov_par} introduces independent non-zero overlap parameters. As we see later in this section, the number of cycles-$6$ can be expressed in terms of the overlap parameters. Thus, the significance of Lemma~\ref{lemma_ind_nz_ov_par} is to reduce the complexity of the discrete optimization problem that specifies the optimum values for the overlap parameters, and consequently, the optimal partitioning.
\vspace{-0.3cm}
\begin{lemma}\label{lemma_ind_nz_ov_par}
The set of independent non-zero overlap parameters $\mathcal{O}_\text{ind}$ is:\vspace{-0.2cm}
\begin{equation}\label{equ_ind_nz_ov_par}
\mathcal{O}_\text{ind}=\{t_{\{i_1,\cdots,i_d\}}|1\leq d\leq \gamma,0\leq i_1,\cdots,i_d<m\gamma,\forall \{i_u,i_v\}\subset\{i_1,\cdots,i_d\}\hspace{0.1cm}i_u\not\equiv i_v \text{ (mod $\gamma$)}\}.\\
\end{equation}
The other overlap parameters not included in $\mathcal{O_\text{ind}}$ are either zero or functions of the overlap parameters in $\mathcal{O_\text{ind}}$. Let $0\leq i_1,\cdots,i_{d_1} < m\gamma$, $m\gamma \leq j_1,\cdots,j_{d_2} < (m+1)\gamma$, and $1\leq(d_1+d_2)\leq \gamma$. Then, $t_{\{i_1,\cdots,i_{d_1},j_1,\cdots,j_{d_2}\}}\in\mathcal{O}\setminus\mathcal{O}_\text{ind}$ is a linear function of the overlap parameters in $\mathcal{O}_\text{ind}$:\vspace{-0.1cm}
\begin{equation}\label{equ_ind_nz_ov_par_relation}
t_{\{i_1,\cdots,i_{d_1},j_1,\cdots,j_{d_2}\}}=\begin{cases}
t_\mathcal{I}+\sum\limits_{\alpha=1}^{d_2}{(-1)^\alpha}\sum\limits_{\substack{\{j_1',\cdots, j_\alpha'\}\subset\mathcal{J}\\ [ x_1,\cdots,x_\alpha ] \in\{0,\cdots,m-1\}^\alpha}}t_{\mathcal{I}\cup\{x_1\gamma+\overline{j_1'},\cdots,x_{\alpha}\gamma+\overline{j_\alpha'}\}},&\mathcal{I}\neq \varnothing,\\
\vspace{0.5cm}
\kappa+\sum\limits_{\alpha=1}^{d_2}{(-1)^\alpha}\sum\limits_{\substack{\{j_1',\cdots, j_\alpha'\}\subset\mathcal{J}\\ [ x_1,\cdots,x_\alpha ] \in\{0,\cdots,m-1\}^\alpha}}t_{\{x_1\gamma+\overline{j_1'},\cdots,x_{\alpha}\gamma+\overline{j_\alpha'}\}},&\mathcal{I}= \varnothing,\vspace{-0.2cm}
\end{cases}
\end{equation}
where $\mathcal{I}=\{i_1,\cdots,i_{d_1}\}$, $\mathcal{J}=\{j_1,\cdots,j_{d_2}\}$, and $\overline{j}=(j\text{ mod }\gamma)$.
\end{lemma}\vspace{-0.4cm}
We note that row $x\gamma+\overline{j}$, where $x\in\{0,\cdots,m-1\}$ and $j\in\mathcal{J}$, belongs to the $x$'th component matrix of $\mathbf{\Pi}_1^{1,p}$, and corresponds to row $j$ of $\mathbf{\Pi}_1^{1,p}$. 
\begin{IEEEproof}
\vspace{-0.1cm}
The proof is given in Appendix A.
\end{IEEEproof}
\begin{example}\label{example_ind_ov_par}\vspace{-0.2cm}
The set of independent non-zero overlap parameters for an SC code with parameters $m=1$ and $\gamma=3$ is:\vspace{-0.1cm}
\begin{equation*}
\mathcal{O}_\text{ind}=\{t_0,t_1,t_2,t_{\{0,1\}},t_{\{0,2\}},t_{\{1,2\}},t_{\{0,1,2\}}\}.\vspace{-0.15cm}
\end{equation*}
According to (\ref{equ_ind_nz_ov_par_relation}), the overlap parameters not in $\mathcal{O}_\text{ind}$ are functions of the $7$ overlap parameters in $\mathcal{O}_\text{ind}$ as follows:
\vspace{0.25cm}

\begin{tabular}{lll}
$t_{3}=\kappa-t_0$,&\hspace{+4.3cm}$t_{\{3,4\}}=\kappa-t_0-t_1+t_{\{0,1\}}$,\\
$t_{4}=\kappa-t_1$,&\hspace{+4.3cm}$t_{\{3,5\}}=\kappa-t_0-t_2+t_{\{0,2\}}$,\\
$t_{5}=\kappa-t_2$,&\hspace{+4.3cm}$t_{\{4,5\}}=\kappa-t_1-t_2+t_{\{1,2\}}$,
\end{tabular}

\begin{tabular}{lll}
$t_{\{0,4\}}=t_0-t_{\{0,1\}}$,\text{\hspace{3cm}}&$t_{\{0,1,5\}}=t_{\{0,1\}}-t_{\{0,1,2\}}$,\\
$t_{\{0,5\}}=t_0-t_{\{0,2\}}$,&$t_{\{0,2,4\}}=t_{\{0,2\}}-t_{\{0,1,2\}}$,\\
$t_{\{1,3\}}=t_1-t_{\{0,1\}}$,&$t_{\{0,4,5\}}=t_0-t_{\{0,1\}}-t_{\{0,2\}}+t_{\{0,1,2\}}$,\\
$t_{\{1,5\}}=t_1-t_{\{1,2\}}$,&$t_{\{1,2,3\}}=t_{\{1,2\}}-t_{\{0,1,2\}}$,\\
$t_{\{2,3\}}=t_2-t_{\{0,2\}}$,&$t_{\{1,3,5\}}=t_1-t_{\{0,1\}}-t_{\{1,2\}}+t_{\{0,1,2\}}$,\\
$t_{\{2,4\}}=t_2-t_{\{1,2\}}$,&$t_{\{2,3,4\}}=t_2-t_{\{0,2\}}-t_{\{1,2\}}+t_{\{0,1,2\}}$,\\
\multicolumn{2}{l}{$t_{\{3,4,5\}}=\kappa-t_0-t_1-t_2+t_{\{0,1\}}+t_{\{0,2\}}+t_{\{1,2\}}-t_{\{0,1,2\}}$.}&
\end{tabular}
\end{example}

Lemma~\ref{lemma_num_in_nz_ov_par} describes the number of elements in $\mathcal{O}_\text{ind}$.\vspace{-0.4cm}
\begin{lemma}\label{lemma_num_in_nz_ov_par}
The number of independent non-zero overlap parameters described in Lemma~\ref{lemma_ind_nz_ov_par} is:
\begin{equation}\label{equ_num_in_nz_ov_par}
\mathcal{N}_\text{ind}=|\mathcal{O}_\text{ind}|=\sum_{d=1}^{\gamma}m^d {{\gamma}\choose{d}}.
\end{equation}
\end{lemma}

\begin{IEEEproof}
The set $\mathcal{O}_\text{ind}$ can be partitioned into $\gamma$ disjoint subsets as follows:
\begin{equation}\label{equ_partitioning_o_ind}
\begin{split}
&\mathcal{O}_\text{ind}=\bigcup_{d=1}^\gamma \mathcal{O}_\text{ind}^d,\\
&\mathcal{O}_\text{ind}^d=\{t_{\{i_1,\cdots,i_d\}}|0\leq i_1,\cdots,i_d<m\gamma,\forall \{i_u,i_v\}\subset\{i_1,\cdots,i_d\}\hspace{0.1cm}i_u\not\equiv i_v \text{ (mod $\gamma$)}\}.
\end{split}
\end{equation}\vspace{+0.1cm}
Consequently,\vspace{-0.1cm}
\begin{equation}\label{equ_partitioning_o_ind_sum}
\mathcal{N}_\text{ind}=\sum_{d=1}^{\gamma}|\mathcal{O}_\text{ind}^k|.
\end{equation}
According to (\ref{equ_partitioning_o_ind}), $|\mathcal{O}_\text{ind}^d|$ is the number of subsets of the set $S=\{0,1,\cdots,m\gamma-1\}$ with size $d$ such that no two elements in the same subset have the same values mod $\gamma$. We first partition the set $S$ as follows:\vspace{-0.2cm}
\begin{equation}\label{equ_partitioning_S}
S=\{0,\gamma,\cdots,(m-1)\gamma\}\cup\{1,\gamma+1,\cdots,(m-1)\gamma+1\}\cup\cdots\cup\{\gamma-1,2\gamma-1,\cdots,m\gamma-1\}.\vspace{-0.2cm}
\end{equation}
All elements of a partition in (\ref{equ_partitioning_S}) have the same value mod $\gamma$. Thus, we have to pick at most one element from each partition to form a subset with the specified characteristics. For choosing $\{i_1,\cdots,i_d\}$ such that all conditions in (\ref{equ_partitioning_o_ind}) are satisfied, we first choose $d$ partitions from $\gamma$ partitions in (\ref{equ_partitioning_S}), and then we choose one element from each partition (all partitions are of size $m$). As a result,  $|\mathcal{O}_\text{ind}^d|=m^d{{\gamma}\choose{d}}$ and $\mathcal{N}_\text{ind}=\sum_{d=1}^\gamma m^d{{\gamma}\choose{d}}$. 
\end{IEEEproof}
\begin{example}\label{example_comparson_size_o_o_ind}\vspace{-0.4cm}
The number of independent non-zero overlap parameters for $\gamma=3$ and $m=1$ is $|\mathcal{O}_{\text{ind}}|=7$, while the number of non-zero overlap parameters is $|\mathcal{O}|=26$ (see Example~\ref{example_nz_ov_par} and \ref{example_ind_ov_par}). This comparison shows the importance of Lemma~\ref{lemma_ind_nz_ov_par} in reducing the size of necessary overlap parameters. This number, as we show later, determines the size of a discrete optimization problem that produces optimal partitioning.\vspace{-0.4cm}
\end{example}

Next, we show that the number of common denominator instances for a protograph of an SC code can be expressed as a function of the independent non-zero overlap parameters, i.e., $\mathcal{O}_\text{ind}$. As we noted, a cycle-$6$ is the common denominator of several problematic objects for CB codes with different column weights over AWGN channels. A cycle-$6$ is formed of three distinct overlaps, and each overlap corresponds to one VN in the graph of the code, see Fig.~5.\vspace{-0.2cm}

\begin{figure}
\centering
\includegraphics[width=0.2\textwidth]{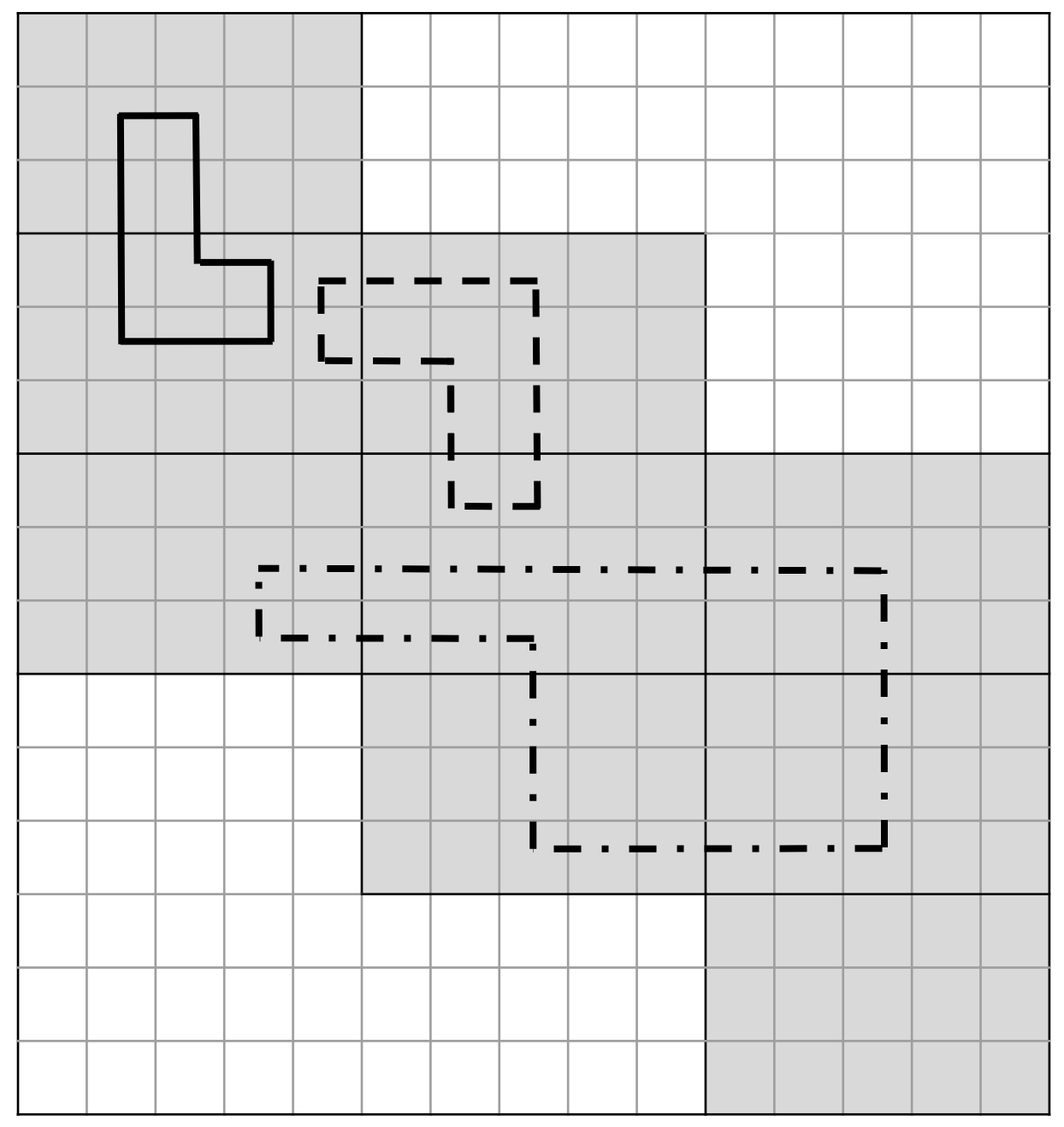}\vspace{-0.5cm}
\caption{Examples of cycles-$6$ on $\mathbf{H}_{SC}^p$ with parameters $\kappa=5$, $\gamma=3$, $m=2$, $L=3$. The cycles with solid lines, dash lines, and dash-dot lines are spanning one, two, and three replicas, respectively. Component matrices are illustrated in gray.\vspace{-0.4cm}}
\vspace{-0.7cm}
\end{figure}
\begin{lemma}\label{lemma_A_B_C}\vspace{-0.15cm}
Consider the protograph of an SC code with parameters $m$, $L$, and $\mathcal{O}$, and let $[x]^+=\max\{x,0\}$. We partition cycles-$6$ with specific CNs into three categories, and enumerate them separately. Three VNs of a cycle-$6$ belong to one, two, or three different replicas. Let the cycle-$6$ starts in $\mathbf{R}_r$ and $c_1=(r-1)\gamma+i_1$, $c_2=(r-1)\gamma+i_2$, and $c_3=(r-1)\gamma+i_3$ be CNs.
\begin{enumerate}
\item The number of cycles-$6$ with all VNs in one replica, say $\mathbf{R}_r$, and CNs $c_1$, $c_2$, and $c_3$, is:
\begin{equation}\label{equ_A}\vspace{-0.1cm}
\begin{split}
\mathcal{A}(t_{\{i_1,i_2,i_3\}}&,t_{\{i_1,i_2\}},t_{\{i_1,i_3\}},t_{\{i_2,i_3\}})=\left(t_{\{i_1,i_2,i_3\}}[t_{\{i_1,i_2,i_3\}}-1]^+[t_{\{i_2,i_3\}}-2]^+\right)\\
+&\left(t_{\{i_1,i_2,i_3\}}(t_{\{i_1,i_3\}}-t_{\{i_1,i_2,i_3\}})[t_{\{i_2,i_3\}}-1]^+\right)\\
+&\left((t_{\{i_1,i_2\}}-t_{\{i_1,i_2,i_3\}})t_{\{i_1,i_2,i_3\}}[t_{\{i_2,i_3\}}-1]^+\right)\\
+&\left((t_{\{i_1,i_2\}}-t_{\{i_1,i_2,i_3\}})(t_{\{i_1,i_3\}}-t_{\{i_1,i_2,i_3\}})t_{\{i_2,i_3\}}\right).
\end{split}
\end{equation}
\item The number of cycles-$6$ with VNs in two replicas, say two VNs in $\mathbf{R}_r$ and one VN in $\mathbf{R}_q$, and CNs $c_1$, $c_2$, and $c_3$, is:\vspace{-0.4cm}
\begin{equation}\label{equ_B}\vspace{-0.1cm}
\begin{split}
\mathcal{B}(t_{\{i_1,i_2,i_3\}}&,t_{\{i_1,i_2\}},t_{\{i_1,i_3\}},t_{\{i_2+(r-q)\gamma,i_3+(r-q)\gamma\}})\\
=&\left(t_{\{i_1,i_2,i_3\}}[t_{\{i_1,i_3\}}-1]^+t_{\{i_2+(r-q)\gamma,i_3+(r-q)\gamma\}}\right)\\
+&\left((t_{\{i_1,i_2\}}-t_{\{i_1,i_2,i_3\}})t_{\{i_1,i_3\}}t_{\{i_2+(r-q)\gamma,i_3+(r-q)\gamma\}}\right),
\end{split}
\end{equation}
where $c_2$ and $c_3$ are the CNs connected via the VN that belongs to $\mathbf{R}_q$.
\item The number of cycles-$6$ with VNs in three replicas, say $\mathbf{R}_r$ and $\mathbf{R}_q$ and $\mathbf{R}_s$ ($r<q<s$), and CNs $c_1$, $c_2$, and $c_3$, is:\vspace{-0.4cm}
\begin{equation}\label{equ_C}\vspace{-0.2cm}
\begin{split}
\mathcal{C}(t_{\{i_1,i_2\}}&,t_{\{i_1+(r-q)\gamma,i_3+(r-q)\gamma\}},t_{\{i_2+(r-s)\gamma,i_3+(r-s)\gamma\}})\\
=&t_{\{i_1,i_2\}}t_{\{i_1+(r-q)\gamma,i_3+(r-q)\gamma\}}t_{\{i_2+(r-s)\gamma,i_3+(r-s)\gamma\}},
\end{split}\vspace{-0.0cm}
\end{equation}
where $c_1$ and $c_2$ are connected via the VN that belongs $\mathbf{R}_r$, $c_1$ and $c_3$ are connected via the VN that belongs to $\mathbf{R}_q$, and $c_2$ and $c_3$ are connected via the VN that belongs to $\mathbf{R}_s$.
\end{enumerate}
\end{lemma}\vspace{-0.3cm}
\begin{IEEEproof}
The proof is given in Appendix B.
\end{IEEEproof}

\vspace{-0.1cm}
\begin{remark}\label{remark_cycle_6_ov_deg_3}\vspace{-0.1cm}
For the enumeration of cycles-$6$, we only need overlap parameters of at most degree three, since the three constituent overlaps of a cycle-$6$ are defined over the three rows involved in the cycle regardless of the column weight. \vspace{-0.2cm}
\end{remark}

Theorem~\ref{theorem_OO} expresses the number of cycles-$6$ in the protograph of an SC code as a function the overlap parameters. We recall that given the independent non-zero overlap parameters, the rest of non-zero overlap parameters can be found using Lemma~\ref{lemma_ind_nz_ov_par}, and $\overline{i}=(i\text{ mod }\gamma)$.\vspace{-0.2cm}
\begin{theorem}\label{theorem_OO}
The number of cycles-$6$ in the protograph of an SC code with parameters $\kappa$, $\gamma$, $m$, $L$, and $\mathcal{O}_{\text{ind}}$ is:
\begin{equation}\label{equ_enumeration_proto}\vspace{-0.2cm}
F=\sum_{k=1}^{m+1} (L-k+1)F_1^k,
\end{equation}
and $F_1^k$, $1\le k \leq (m+1)$, is formulated as follows:\\
\vspace{-0.1cm}
\begin{equation}\label{equ_enumeration_proto_Fk}
\begin{split}
F_1^1&=\sum_{\substack{
\{i_1,i_2,i_3\}\subset{\{0,\cdots,(m+1)\gamma-1\}}\\
\overline{i_1}\neq\overline{i_2}\hspace{0.1cm},\overline{i_1}\neq\overline{i_3},\hspace{0.1cm}\overline{i_2}\neq\overline{i_3}}}
\mathcal{A}(t_{\{i_1,i_2,i_3\}},t_{\{i_1,i_2\}},t_{\{i_1,i_3\}},t_{\{i_2,i_3\}}).\\
F_1^2&=\sum_{\substack{
i_1\in{\{0,\cdots,(m+1)\gamma-1\}}\\
\{i_2,i_3\}\subset{\{\gamma,\cdots,(m+1)\gamma-1\}}\\
\overline{i_1}\neq\overline{i_2}\hspace{0.1cm},\overline{i_1}\neq\overline{i_3},\hspace{0.1cm}\overline{i_2}\neq\overline{i_3}}}
\mathcal{B}(t_{\{i_1,i_2,i_3\}},t_{\{i_1,i_2\}},t_{\{i_1,i_3\}},t_{\{i_2-\gamma,i_3-\gamma\}})\\
&+\sum_{\substack{
i_1\in{\{0,\cdots,(m+1)\gamma-1\}}\\
\{i_2,i_3\}\subset{\{0,\cdots,m\gamma-1\}}\\
\overline{i_1}\neq\overline{i_2}\hspace{0.1cm},\overline{i_1}\neq\overline{i_3},\hspace{0.1cm}\overline{i_2}\neq\overline{i_3}}}
\mathcal{B}(t_{\{i_1,i_2,i_3\}},t_{\{i_1,i_2\}},t_{\{i_1,i_3\}},t_{\{i_2+\gamma,i_3+\gamma\}}).\\
F_1^{k\ge3}&=\sum_{\substack{
i_1\in{\{0,\cdots,(m+1)\gamma-1\}}\\
\{i_2,i_3\}\subset{\{(k-1)\gamma,\cdots,(m+1)\gamma-1\}}\\
\overline{i_1}\neq\overline{i_2}\hspace{0.1cm},\overline{i_1}\neq\overline{i_3},\hspace{0.1cm}\overline{i_2}\neq\overline{i_3}}}
\mathcal{B}(t_{\{i_1,i_2,i_3\}},t_{\{i_1,i_2\}},t_{\{i_1,i_3\}},t_{\{i_2+(1-k)\gamma,i_3+(1-k)\gamma\}})\\
&+\sum_{\substack{
i_1\in{\{0,\cdots,(m+1)\gamma-1\}}\\
\{i_2,i_3\}\subset{\{0,\cdots,(m-k+2)\gamma-1\}}\\
\overline{i_1}\neq\overline{i_2}\hspace{0.1cm},\overline{i_1}\neq\overline{i_3},\hspace{0.1cm}\overline{i_2}\neq\overline{i_3}}}
\mathcal{B}(t_{\{i_1,i_2,i_3\}},t_{\{i_1,i_2\}},t_{\{i_1,i_3\}},t_{\{i_2+(k-1)\gamma,i_3+(k-1)\gamma\}})\\
&+\sum_{q=2}^{k-1}\sum_{\substack{
i_1\in{\{(q-1)\gamma,\cdots,(m+1)\gamma-1\}}\\
i_2\in{\{(k-1)\gamma,\cdots,(m+1)\gamma-1\}}\\
i_3\in{\{(k-1)\gamma,\cdots,(m+q)\gamma-1\}}\\
\overline{i_1}\neq\overline{i_2}\hspace{0.1cm},\overline{i_1}\neq\overline{i_3},\hspace{0.1cm}\overline{i_2}\neq\overline{i_3}}}
\mathcal{C}(t_{\{i_1,i_2\}},t_{\{i_1+(1-q)\gamma,i_3+(1-q)\gamma\}},t_{\{i_2+(1-k)\gamma,i_3+(1-k)\gamma\}})\\
\end{split}
\end{equation}
The functions $\mathcal{A}$, $\mathcal{B}$, and $\mathcal{C}$ are defined in Lemma~\ref{lemma_A_B_C}.
\end{theorem}\vspace{-0.3cm}
\begin{IEEEproof}
The proof is given in Appendix C.
\end{IEEEproof}
Now, define $F^*$ to be the minimum number of cycles-$6$ for $\mathbf{H}_{SC}^p$. Thus, our discrete optimization problem is formulated as follows:
\vspace{-0.2cm}
\begin{equation}\label{equ_F_opt}
F^*=\min_{\mathcal{O}_\text{ind}}F.
\end{equation}
The constraints of our optimization problem are the conditions under which the overlap parameters are valid.
\begin{example}\label{example_constraints}
\vspace{-0.3cm}
For an SC code with parameters $\gamma=3$, and $m=1$, the constraints of the optimization problem in (\ref{equ_F_opt}) are the conditions under which the independent non-zero overlap parameters are valid. Thus, these constraints on the $7$ parameters are (see Example~\ref{example_ind_ov_par}):\\
\vspace{-0.1cm}
\begin{equation}\label{equ_ex_ind_ov_par}
\begin{array}{ll}
0\leq t_0 \leq \kappa,&0\leq t_{\{0,1\}} \leq t_0,\\
t_{\{0,1\}}\leq t_1 \leq \kappa-t_0+t_{\{0,1\}},&0\leq t_{\{0,1,2\}} \leq t_{\{0,1\}},\\
t_{\{0,1,2\}}\leq t_{\{0,2\}}\leq t_0-t_{\{0,1\}}+t_{\{0,1,2\}},&t_{\{0,1,2\}}\leq t_{\{1,2\}}\leq t_1-t_{\{0,1\}}+t_{\{0,1,2\}},\\
\multicolumn{2}{l}{t_{\{0,2\}}+t_{\{1,2\}}-t_{\{0,1,2\}}\leq t_2 \leq \kappa - t_0 - t_1 + t_{\{0,1\}} + t_{\{0,2\}}+t_{\{1,2\}}-t_{\{0,1,2\}},}\\
\lfloor\frac{3\kappa}{2}\rfloor\leq t_0+t_1+t_2 \leq \lceil\frac{3\kappa}{2}\rceil.&
\end{array}
\end{equation}
The last constraint in (\ref{equ_ex_ind_ov_par}) guarantees balanced partitioning between $\mathbf{H}_0^p$ and $\mathbf{H}_1^p$.\vspace{-0.2cm}
\end{example}
Consider an underlying block code with parameters $\kappa$ and $\gamma$. Each circulant of the parity-check matrix $\mathbf{H}$ of this code can belong to any of $(m+1)$ components, leading to $(m+1)^{\kappa\gamma}$ possible s. We want to choose a partitioning that results in the least number of cycles-$6$ in the protograph of an SC code. Considering all possible ways of partitioning in a brute force fashion to find the optimal one is not practical. We reduced the problem of finding an optimum partitioning for the protograph of an SC code, which results in the minimum number of cycles-$6$, into an optimization problem over a number of overlap parameters described in (\ref{equ_ind_nz_ov_par}). Example~\ref{example_OO} summarizes all necessary steps for finding the optimal partitioning of an SC code with parameters $\kappa$, $\gamma=3$, $m=1$, and $L$.\vspace{-0.2cm}
\begin{example}\label{example_OO}
Using Theorem~\ref{theorem_OO}, the number of cycles-$6$ in the protograph of an SC code with parameters $\kappa$, $\gamma=3$, $m=1$, and $L$ is described in terms of the $7$ overlap parameters in $\mathcal{O}_\text{ind}=\{t_0,t_1,t_2,t_{\{0,1\}},t_{\{0,2\}},t_{\{1,2\}},t_{\{0,1,2\}}\}$:
\begin{equation*}
F=LF_1^1+(L-1)F_1^2,
\end{equation*}
where $F_1^1$ and $F_1^2$ are:
\begin{equation}\label{equ_Fk_example_1}
\begin{split}
F_1^1&= \mathcal{A}(t_{\{0,1,2\}},t_{\{0,1\}},t_{\{0,2\}},t_{\{1,2\}})+ \mathcal{A}(t_{\{0,1,5\}},t_{\{0,1\}},t_{\{0,5\}},t_{\{1,5\}})+ \mathcal{A}(t_{\{0,2,4\}},t_{\{0,2\}},t_{\{0,4\}},t_{\{2,4\}})\\
&+ \mathcal{A}(t_{\{0,4,5\}},t_{\{0,4\}},t_{\{0,5\}},t_{\{4,5\}})+ \mathcal{A}(t_{\{1,2,3\}},t_{\{1,2\}},t_{\{1,3\}},t_{\{2,3\}})+ \mathcal{A}(t_{\{1,3,5\}},t_{\{1,3\}},t_{\{1,5\}},t_{\{3,5\}})\\
&+ \mathcal{A}(t_{\{2,3,4\}},t_{\{2,3\}},t_{\{2,4\}},t_{\{3,4\}})+ \mathcal{A}(t_{\{3,4,5\}},t_{\{3,4\}},t_{\{3,5\}},t_{\{4,5\}}),
\end{split}
\end{equation}
\begin{equation}\label{equ_Fk_example_2}
\begin{split}
F_1^2&=\mathcal{B}(t_{\{0,4,5\}},t_{\{0,4\}},t_{\{0,5\}},t_{\{1,2\}})+\mathcal{B}(t_{\{3,4,5\}},t_{\{3,4\}},t_{\{3,5\}},t_{\{1,2\}})+\mathcal{B}(t_{\{1,3,5\}},t_{\{1,3\}},t_{\{1,5\}},t_{\{0,2\}})\\
&+\mathcal{B}(t_{\{3,4,5\}},t_{\{3,4\}},t_{\{4,5\}},t_{\{0,2\}})+\mathcal{B}(t_{\{2,3,4\}},t_{\{2,3\}},t_{\{2,4\}},t_{\{0,1\}})+\mathcal{B}(t_{\{3,4,5\}},t_{\{3,5\}},t_{\{4,5\}},t_{\{0,1\}})\\
&+\mathcal{B}(t_{\{0,1,2\}},t_{\{0,1\}},t_{\{0,2\}},t_{\{4,5\}})+\mathcal{B}(t_{\{1,2,3\}},t_{\{1,3\}},t_{\{2,3\}},t_{\{4,5\}})+\mathcal{B}(t_{\{0,1,2\}},t_{\{0,1\}},t_{\{1,2\}},t_{\{3,5\}})\\
&+\mathcal{B}(t_{\{0,2,4\}},t_{\{0,4\}},t_{\{2,4\}},t_{\{3,5\}})+\mathcal{B}(t_{\{0,1,2\}},t_{\{0,2\}},t_{\{1,2\}},t_{\{3,4\}})+\mathcal{B}(t_{\{0,1,5\}},t_{\{0,5\}},t_{\{1,5\}},t_{\{3,4\}}).
\end{split}
\end{equation}
We note that all the overlap parameters in (\ref{equ_Fk_example_1}) and (\ref{equ_Fk_example_2}) are linear combinations of the $7$ independent non-zero overlap parameters in $\mathcal{O}_{\text{ind}}$, see Example~\ref{example_ind_ov_par}. The functions $\mathcal{A}$ and $\mathcal{B}$ are defined in Lemma~\ref{lemma_A_B_C}. Our discrete optimization problem is formulated as follows:
\begin{equation}\label{equ_OO_example}
F^*=\min_{t_0,t_1,t_2,t_{\{0,1\}},t_{\{0,2\}},t_{\{1,2\}},t_{\{0,1,2\}}}F.
\end{equation}
The constraints of the optimization problem are found in Example~\ref{example_constraints}. The solution of our optimization problem is not unique. However, since all the solutions result in the same number of OO partitioning choices and the same $F^*$, we work with one of these solutions, and call it an optimal vector, $\mathbf{t}^* = \left[t_0^*,t_1^*,t_2^*,t_{\{0,1\}}^*,t_{\{0,2\}}^*,t_{\{1,2\}}^*,t_{\{0,1,2\}}^*\right]$.
\end{example}
\subsection{Circulant Power Optimization}
After picking an optimal vector $\mathbf{t}^* $ to partition $\mathbf{H}^{p}$ and construct $\mathbf{H}^{p}_{SC}$, we run our heuristic CPO to further reduce the number of $(3, 3(\gamma-2))$ ASs/TSs (the common denominator instances) in the graph of $\mathbf{H}_{SC}$ with column weight $\gamma$. Recall that in codes that have no cycle-$4$, a cycle-$6$ is a $(3, 3(\gamma-2))$ AS/TS. The steps of the CPO are:
\begin{enumerate}
\item Initially, assign circulant powers as in AB codes, i.e., $f_{i,j} = ij$, $0 \leq i \leq \gamma-1$ and $0 \leq j \leq \kappa-1$, to all the $\gamma \kappa$ $1$'s in $\mathbf{H}^{p}$ (results in no cycle-$4$ in $\mathbf{H}$ and $\mathbf{H}_{SC}$).
\item Construct $\mathbf{\Pi}^{\chi,p}_1$, which contains $\chi = m+1$ replicas and has the size $(2m+1) \gamma \times (m+1) \kappa$ (see Definition~\ref{def_pi}), using $\mathbf{H}^{p}$ and $\mathbf{t}^*$. (Recall that the VNs of a cycle-$6$ span at most $\chi = m+1$ consecutive replicas.) Circulant powers of the $1$'s in $\mathbf{\Pi}^{\chi,p}_1$ are copied from the $1$'s in $\mathbf{H}^{p}$.
\item Define a counting variable $\theta_{i,j}$, $0 \leq i \leq \gamma-1$ and $0 \leq j \leq \kappa-1$, for each of the $1$'s in $\mathbf{H}^{p}$. Define another counting variable $\theta'_{i',j'}$, $0 \leq i' \leq (2m+1)\gamma-1$ and $0 \leq j' \leq (m+1)\kappa-1$, for each of the elements in $\mathbf{\Pi}^{\chi,p}_1$. Initialize all the variables in this step with zeros. Notice that only $(m+1)\gamma \kappa$ counting variables of the form $\theta'_{i',j'}$ are associated with $1$'s in $\mathbf{\Pi}^{\chi,p}_1$ (the remaining counting variables will remain zeros). 
\item Locate all the cycles-$4$ and cycles-$6$ in $\mathbf{\Pi}^{\chi,p}_1$.
\item Specify the cycles-$6$ in $\mathbf{\Pi}^{\chi,p}_1$ that have (\ref{equ_CP}) satisfied, and call them \textit{active cycles}. Let $F^{k,a}_1$, $k \in \{1, \dots, m+1\}$, be the number of active cycles starting at the first replica and having their VNs spanning $k$ consecutive replicas in $\mathbf{\Pi}^{\chi,p}_1$. Thus, from (\ref{equ_enumeration}), the number of active cycles having their VNs spanning $k$ consecutive replicas in $\mathbf{\Pi}^{\chi,p}_1$ is $(m-k+2)F^{k,a}_1$. (For example, for $k=1$, $(m+1)F^{1,a}_1$ is the number of active cycles having their VNs spanning one replica in $\mathbf{\Pi}^{\chi,p}_1$.)
\item Compute the number of $(3, 3(\gamma-2))$ ASs/TSs in $\mathbf{H}_{SC}$ using the following formula (see (\ref{equ_enumeration}) and (\ref{equ_CP})):
\vspace{-0.2em}\begin{equation}\label{equ_F_SC}
F_{SC} = \sum_{k=1}^{m+1} \left ( (L-k+1)F^{k,a}_1 \right ) p. \vspace{-0.2em}
\end{equation}
\item Count the number of active cycles each $1$ in $\mathbf{\Pi}^{\chi,p}_1$ is involved in. Assign weight $w_k = (m+1)/(m-k+2)$ to the number of active cycles having their VNs spanning $k$ consecutive replicas in $\mathbf{\Pi}^{\chi,p}_1$ (see Remark~\ref{step_7_cpo} for more clarification). Store the weighted count associated with each $1$ in $\mathbf{\Pi}^{\chi,p}_1$, which is indexed by $(i', j')$, in $\theta'_{i',j'}$. (For example, for $k=m+1$, the weight of the number of active cycles having their VNs spanning $m+1$ consecutive replicas is $(m+1)$.)
\item Calculate the counting variables $\theta_{i,j}$, $\forall i,j$, associated with the $1$'s in $\mathbf{H}^{p}$ from the counting variables $\theta'_{i',j'}$ associated with the $1$'s in $\mathbf{\Pi}^{\chi,p}_1$ (computed in step 7) using the following formula:
\vspace{-0.2em}\begin{equation}
\theta_{i,j} = \sum_{i': i'_{\gamma}=i} \sum_{\substack{j': j'_{\kappa}=j \\ \mathbf{\Pi}^{\chi,p}_1 [i'][j'] \neq 0}} \theta'_{i',j'}, \vspace{-0.2em}
\end{equation}
where $i'_{\gamma} = (i' \mod \gamma)$ and $j'_{\kappa} = (j' \mod \kappa)$. Sort these $\gamma \kappa$ $1$'s of $\mathbf{H}^{p}$ in a list descendingly according to the counts in $\theta_{i,j}$, $\forall i,j$.
\item Pick a subset of $1$'s from the top of this list, and change the circulant powers associated with them.
\item Using these interim new powers, do steps 5 and 6.
\item If $F_{SC}$ is reduced while maintaining no cycle-$4$ in $\mathbf{H}_{SC}$, update $F_{SC}$ and the circulant powers, then go to step 7. Otherwise, return to step 9 to pick a different set of circulant powers or/and a different subset of $1$'s (from the $1$'s in $\mathbf{H}^{p}$).
\item Iterate until the target $F_{SC}$ (set by the designer) is achieved, or the CPO gain in $F_{SC}$ approaches zero.
\end{enumerate}
Note that step 9 is performed heuristically.

\vspace{-0.3cm}
\begin{remark}\label{step_7_cpo}
A cycle-$6$ starting at the first replica and spanning $k$ consecutive replicas is repeated exactly $(m-k+2)$ times, $\forall k \in \{1, \dots, m+1\}$, in $\mathbf{\Pi}^{\chi,p}_1$. Here, we assume that $L >> \chi -1 = m$, which is the case for practical settings. Consequently, cycles spanning $k$ consecutive replicas, $\forall k \in \{1, \dots, m+1\}$, should be treated the same way in steps 7 and 8 in the CPO algorithm; because they have approximately the same contribution to $F_{SC}$ in (\ref{equ_F_SC}). In order to perform this task, a cycle spanning $k$ consecutive replicas, which involves a $1$ in $\mathbf{\Pi}^{\chi,p}_1$ indexed by $(i',j')$, is counted exactly $(m-k+2)w_k = (m+1)$ times in $\theta'_{i',j'}$ (and in the corresponding $\theta_{i,j}$), leading to the weighting factors used in steps 7 and 8.
\end{remark}

\begin{remark}
\vspace{-0.2cm}
A $(3, b')$ configuration, where $b' <= 3(\gamma -2)$, in the protograph of the SC code can result in $(3, 3(\gamma -2))$ ASs/TSs in the final (lifted) graph depending on the circulant power arrangement. That is the reason why we enumerate all cycles-$6$ (even those that are $(3, b')$ configurations, $b' < 3(\gamma -2)$) in the protograph, and minimize their number. Note that typically the protograph does have cycles-$4$. 
\end{remark}

\begin{example}\label{example_gamma_4}\vspace{-0.2cm}
Suppose we want to design an SC code with $\kappa=p=7$, $\gamma=4$, $m=1$, and $L=30$ using OO-CPO (OO partitioning and CPO algorithm). Solving the optimization problem in (\ref{equ_F_opt}) yields the following optimal overlap parameters:\vspace{-0.2cm}
\begin{equation*}\vspace{-0.1cm}
\begin{split}
\mathbf{t}^*&=\left[t_0^*,t_1^*,t_2^*,t_3^*,t_{\{0,1\}}^*,t_{\{0,2\}}^*,t_{\{0,3\}}^*,t_{\{1,2\}}^*,t_{\{1,3\}}^*,t_{\{2,3\}}^*,t_{\{0,1,2\}}^*,t_{\{0,1,3\}}^*,t_{\{0,2,3\}}^*,t_{\{1,2,3\}}^*\right]\\
&=[3,4,3,4,0,1,2,2,2,0,0,0,0,0],
\end{split}
\end{equation*}
which results in $F^*=4,680$ cycles-$6$ in the graph of $\mathbf{H}_{SC}^p$. Fig~6(a) shows how the partitioning is applied on $\mathbf{H}^p$ (or $\mathbf{H}$). Next, we apply the CPO algorithm to obtain the parity-check matrix $\mathbf{H}_{SC}$. Fig~6(b) shows the circulant power arrangement of $\mathbf{H}_{SC}$ using CPO algorithm. We note that the uncoupled case with AB circulant power arrangement ($\mathbf{H}_0=\mathbf{H}$, $\mathbf{H}_1=\mathbf{0}$, and $f_{i,j}=ij$) results in $\mathbf{H}_{SC}$ with $35,280$ cycles-$6$, OO partitioning with AB circulant power arrangement (only stage~1) results in $\mathbf{H}_{SC}$ with $5,747$ cycles-$6$, and the OO-CPO approach (both stages) results in $\mathbf{H}_{SC}$ with $2,870$ cycles-$6$.\vspace{-0.25cm}

\begin{figure}
\centering
\begin{tabular}{cc}
\vspace{0.07cm}\includegraphics[width=0.2\textwidth]{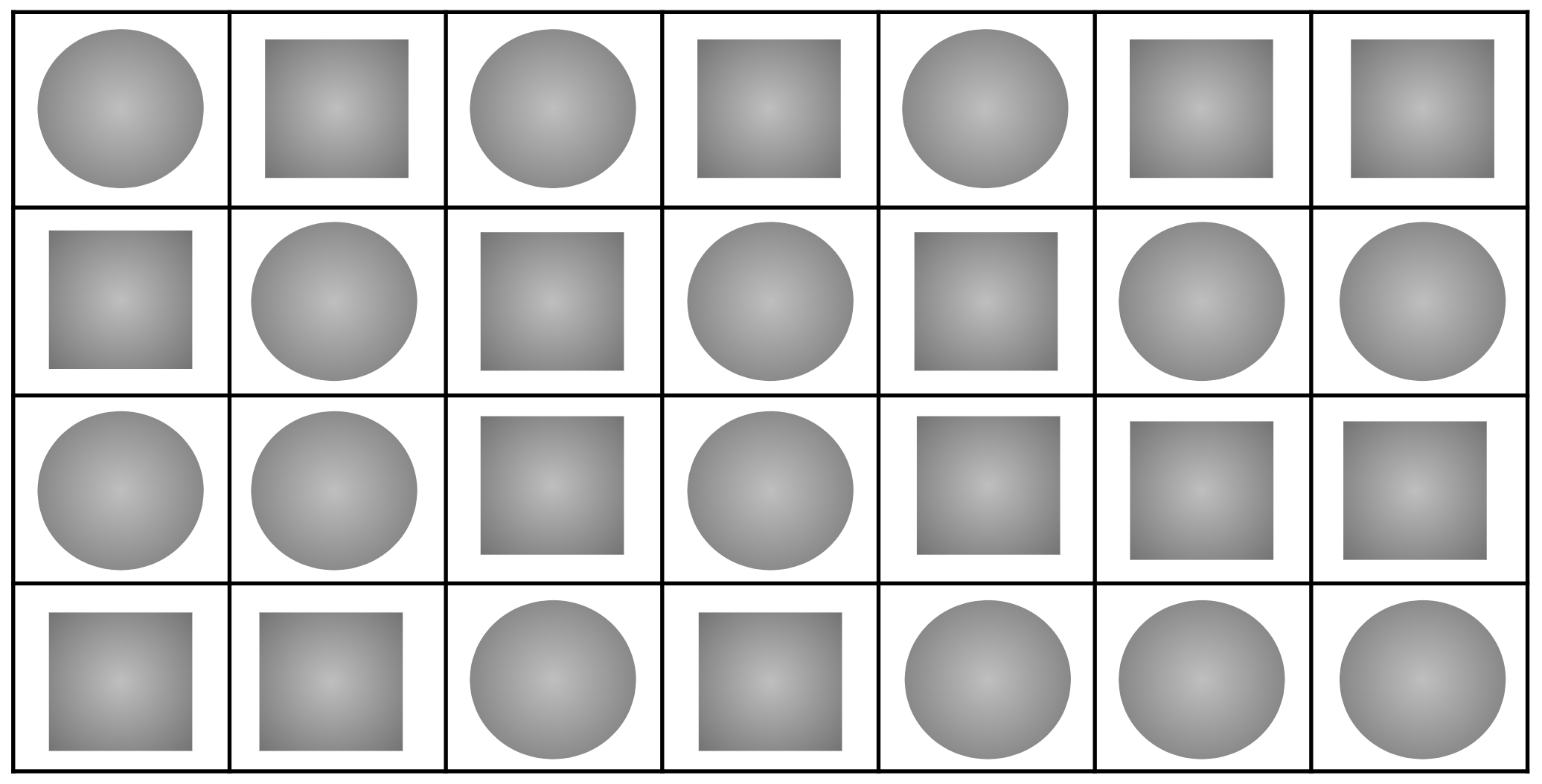}&\includegraphics[width=0.2\textwidth]{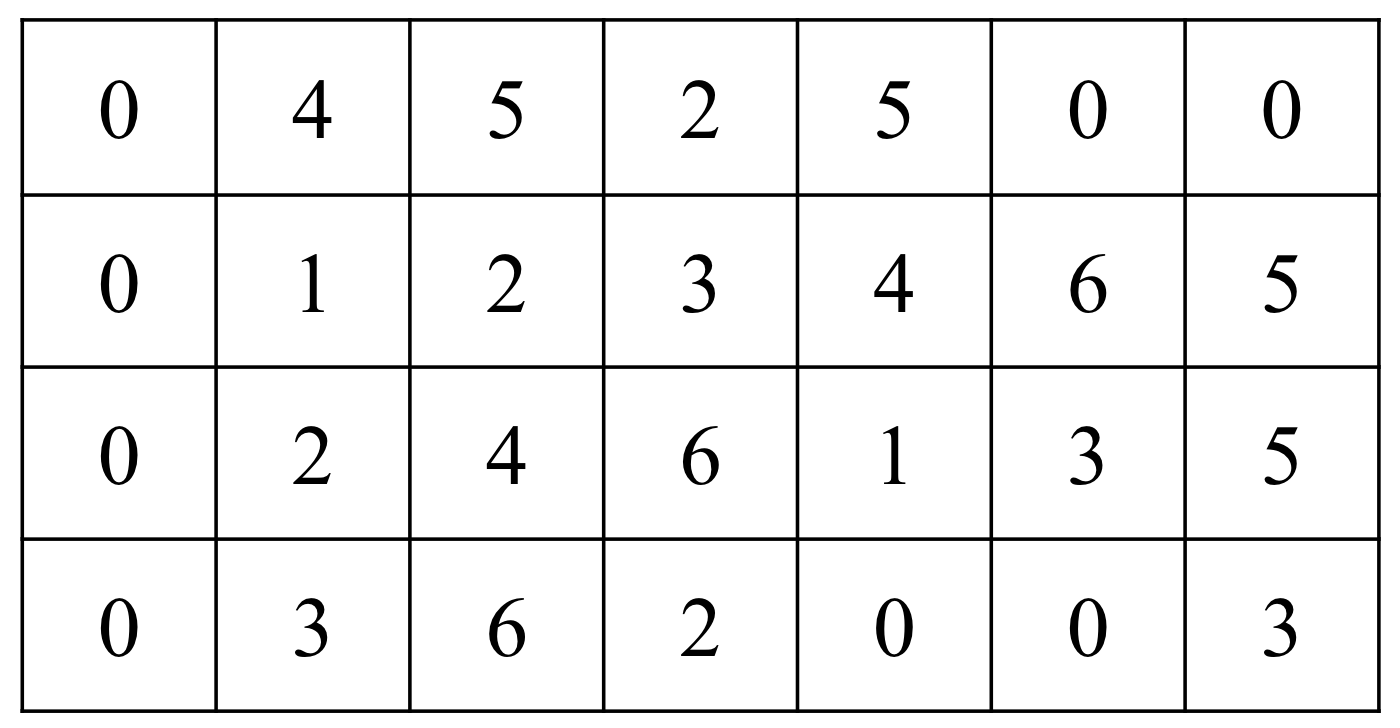}\vspace{-0.5cm}\\
(a)&(b)\vspace{-0.6cm}
\end{tabular}
\caption{(a) The OO partitioning of $\mathbf{H}^p$ (or $\mathbf{H}$) of the SC code in Example~\ref{example_gamma_4}. Entries with circles (resp., squares) are assigned to $\mathbf{H}_0^p$ (resp., $\mathbf{H}_1^p$); (b) The circulant power arrangement for the circulants in $\mathbf{H}^p$.}\vspace{-0.95cm}
\end{figure}
\vspace{-0.3cm}
\end{example}
\section{\vspace{-0.2cm}Simulation Results}

In this section, we compare the error floor performance of SC codes constructed by the  OO-CPO framework with uncoupled block codes and SC codes constructed by the previous method of partitioning by cutting vectors \cite{MitchellISIT2014,AmiriTCOM2016}. First, we demonstrate the reduction of the number of detrimental objects achieved by the OO-CPO framework via Table~I. Then, we show the performance gain via BER curves provided in Figs.~8 and 9. In our simulations, we consider AWGN channels, and SC codes with memories $m\in\{1,2\}$ and $\gamma\in\{3,4\}$.

All the codes we simulated are binary CB-SC codes with $\kappa=p=17$, $L=30$, length $8,670$ bits, and constructed by different methods. Uncoupled~Codes~1 and 2 are uncoupled AB codes (SC codes with $m=0$, $\mathbf{H}_0=\mathbf{H}$, and $f_{i,j}=ij$), and have $\gamma=3$ and $4$, respectively. SC~Codes~1-3 have $m=1$ and $\gamma=3$, SC~Codes~4-6 have $m=1$ and $\gamma=4$, and SC~Codes~7 has $m=2$ and $\gamma=3$. SC~Codes~1 and 4 are constructed by the optimal cutting vectors $\boldsymbol{\zeta}=[4,9,13]$ and $[3,7,11,15]$, respectively, and AB circulant powers \cite{AmiriTCOM2016}. SC~Codes~2 and 5 are constructed by OO partitioning and AB circulant powers. SC~Codes~3, 6, and 7 are constructed by the OO-CPO framework (see Fig.~7).

\begin{figure}
\centering
\begin{tabular}{cc}
\includegraphics[width=0.45\textwidth]{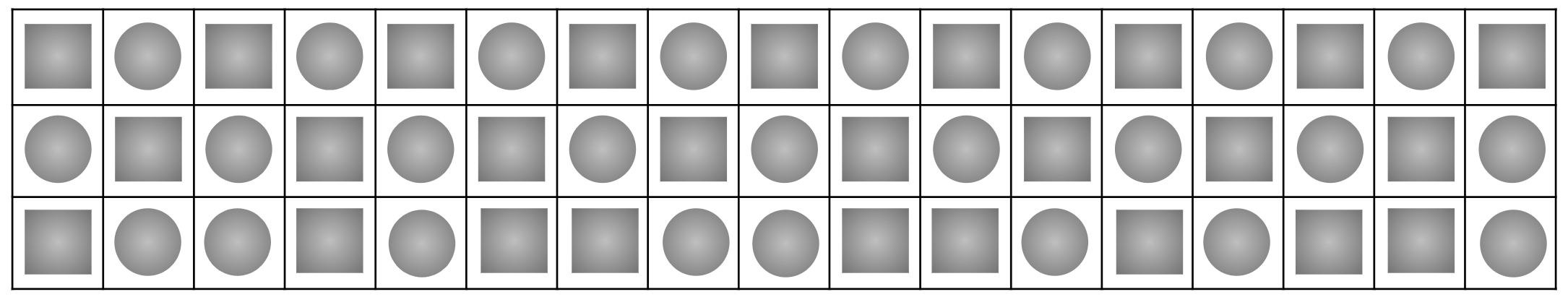}&\includegraphics[width=0.45\textwidth]{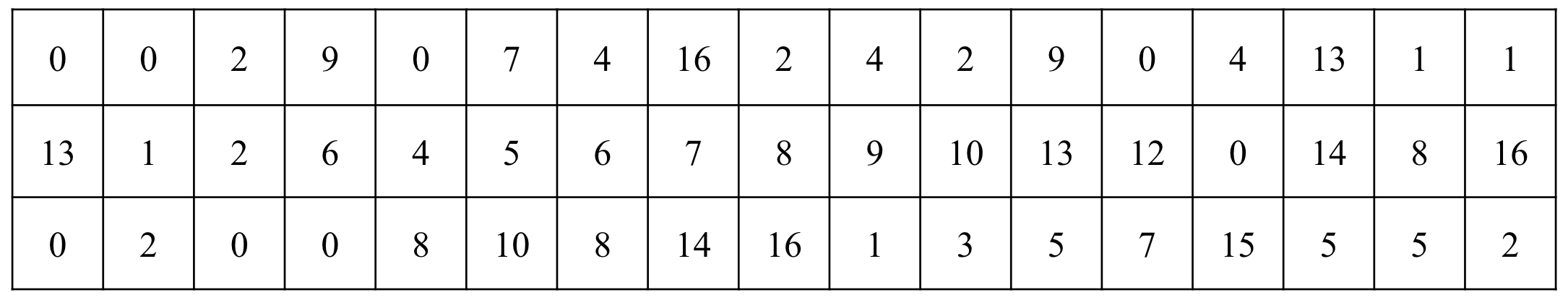}\vspace{-0.55cm}\\
\multicolumn{2}{c}{(a)}\\
\includegraphics[width=0.45\textwidth]{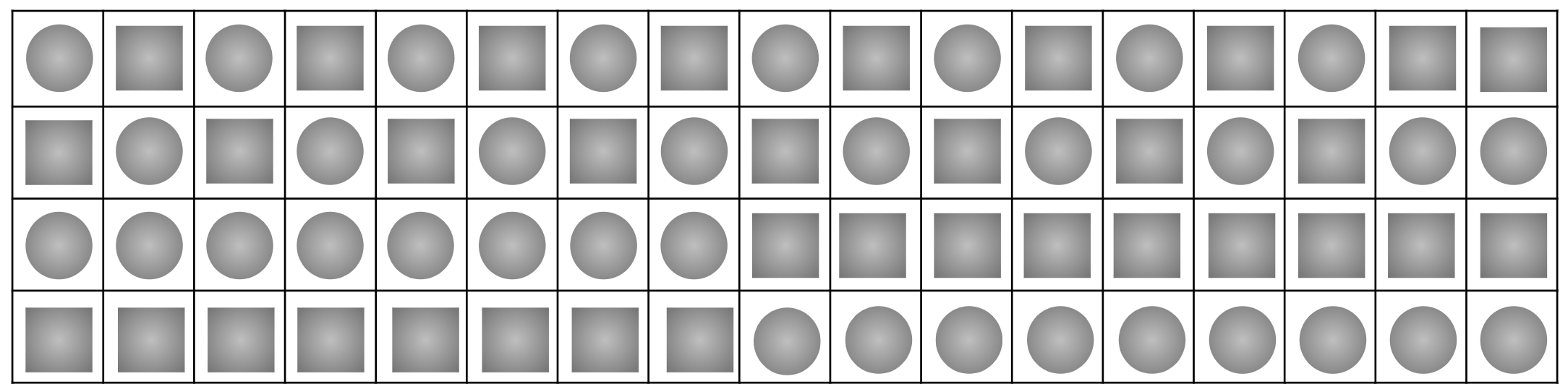}&\includegraphics[width=0.45\textwidth]{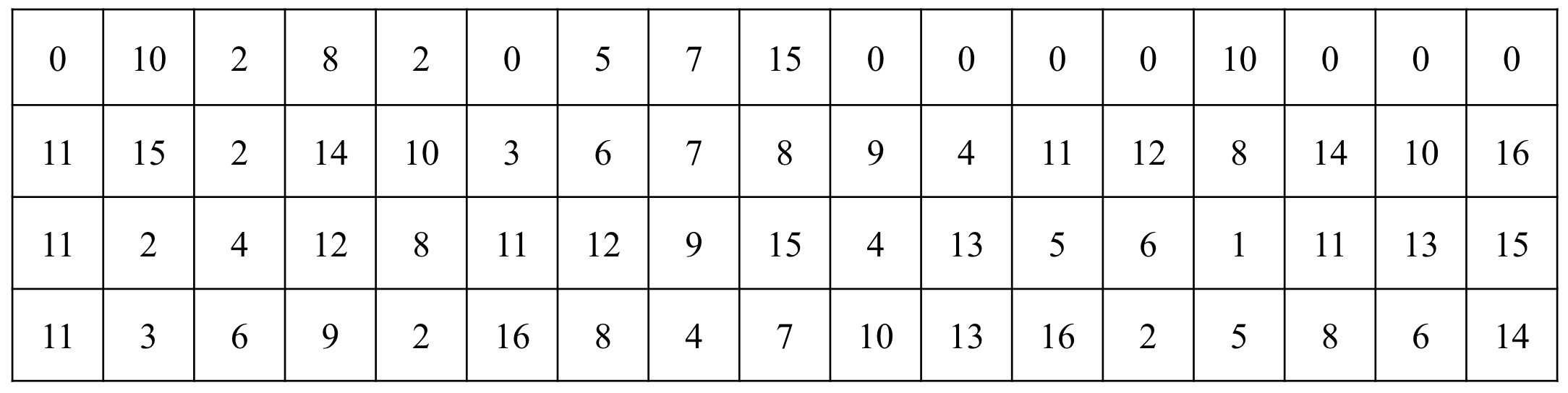}\vspace{-0.55cm}\\
\multicolumn{2}{c}{(b)}\\
\includegraphics[width=0.45\textwidth]{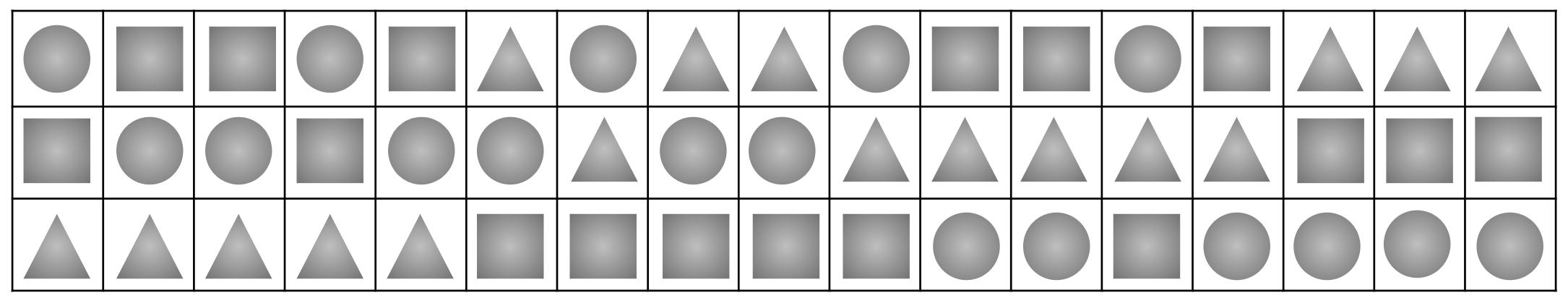}&\includegraphics[width=0.45\textwidth]{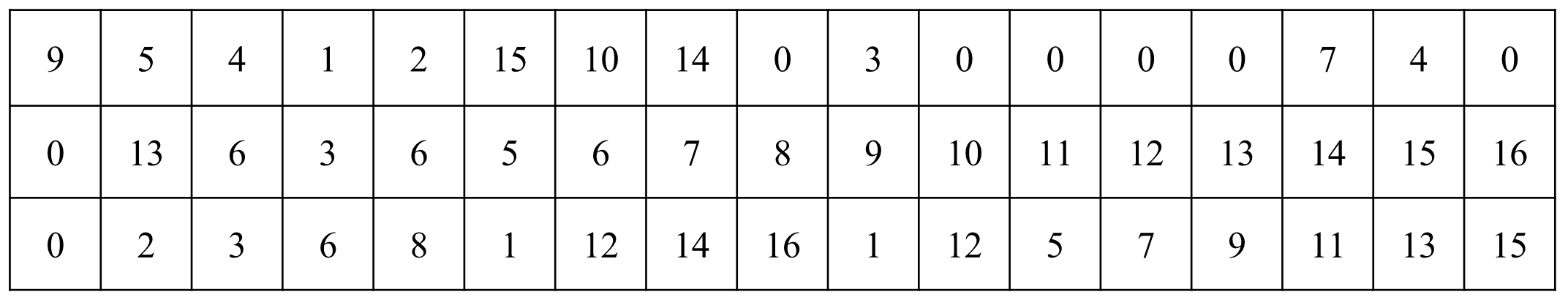}\vspace{-0.55cm}\\
\multicolumn{2}{c}{(c)}\\\vspace{-1.55cm}\\
\end{tabular}
\caption{OO partitioning (left panel) and circulant power arrangement obtained by CPO (right panel) for SC~Codes~3, 6, and 7. Subfigures are organized as follows: (a) SC~Codes~3; (b) SC~Codes~6; (c) SC~Codes~7.}
\vspace{-1.15cm}
\end{figure}

Table~I shows the number of cycles $6$ for Uncoupled~Codes~1-2 and SC~Codes~1-7. According to our results, an SC code constructed by the optimal cutting vector, achieves about $57\%$ reduction in the number of cycles-$6$ ASs compared to the uncoupled case for $m=1$ and $\gamma\in\{3,4\}$. This reduction becomes $80\%$ when the memory is $m=2$. When we apply our new OO-CPO approach for $m=1$ and $\gamma\in\{3,4\}$, the reduction in the population of cycles-$6$ compared to the uncoupled case reaches up to $89\%$, which is even better than the partitioning with optimal cutting vectors with $m=2$. For the OO-CPO approach with $m=2$ and $\gamma=3$, the number of cycles-$6$ (the common denominator instances) is $0$ which means all cycles-$6$ are removed using our approach, and the girth of the code becomes $8$. We present the results for $\kappa=p=17$ in this section; however, our simulation results (which we do not report here because of space limitation) show similar gains for $p=7$, $p=11$, and $p=13$ as well.

\begin{table}[ht]
\caption{\vspace{-0.2cm}Comparison of the population of cycles-$6$ for SC codes with $p=\kappa=17$, $L=30$, and different construction schemes.}
\centering
 \begin{tabular}{|l|c|c|c|} 
 \hline
Construction scheme&$m$&$\gamma$&No. cycles $6$\\
 \hline
 \hline
  \multirow{2}{10em}{Uncoupled case} & \multirow{2}{1em}{--} & $3$ & $138,720$ \\
 \hhline{~~--}
AB circulant arrangement && $4$ & $554,880$\\
 \hline
 \multirow{2}{13em}{Cutting vector partitioning} & $1$ & $3$ & $59,024$\\
 \hhline{~---}
 AB circulant arrangement & $2$ & $3$ & $27,880$\\
  \hhline{~---}
 & $1$ & $4$ & $238,697$\\
 \hline
 \multirow{2}{13em}{Optimal overlap partitioning}  & $1$ & $3$ & $14,960$\\
 \hhline{~---}
 Circulant power optimization& $2$ & $3$ & $0$\\
  \hhline{~---}
 & $1$ & $4$ & $91,494$\\
 \hline
 \end{tabular}\vspace{-1.0cm}
\end{table}


Next, we compare the performance of SC codes constructed using different methods over AWGN channels. Fig~8(a) shows the BER curves in the error floor area for Uncoupled~Codes~1 and SC~Codes~1-3. All these codes have $\kappa=p=17$, $\gamma=3$, $m=1$, and $L=30$. SC~Code~1 is constructed using the cutting vector partitioning, SC~Code~2 is constructed by employing only the first stage of our framework, and SC~Code~3 is constructed by applying the two stages. The figure demonstrates that our method outperforms the cutting vector scheme by more than $2$ orders of magnitude, and that each stage of the framework is necessary to achieve this improvement.

Fig~8(b) shows similar findings for $\gamma=4$ SC codes. Based on our results, the performance improvement of our two-stage framework relative to the cutting vector scheme is more pronounced when we increase the column weight. Additionally, SC~Code~6 constructed by the OO-CPO framework achieves nearly $5$ orders of magnitude performance improvement in the early error floor area and up to $1.5$~dB SNR gain compared to the uncoupled setting.

\begin{figure}
\centering
\begin{tabular}{cc}
\includegraphics[width=0.4\textwidth]{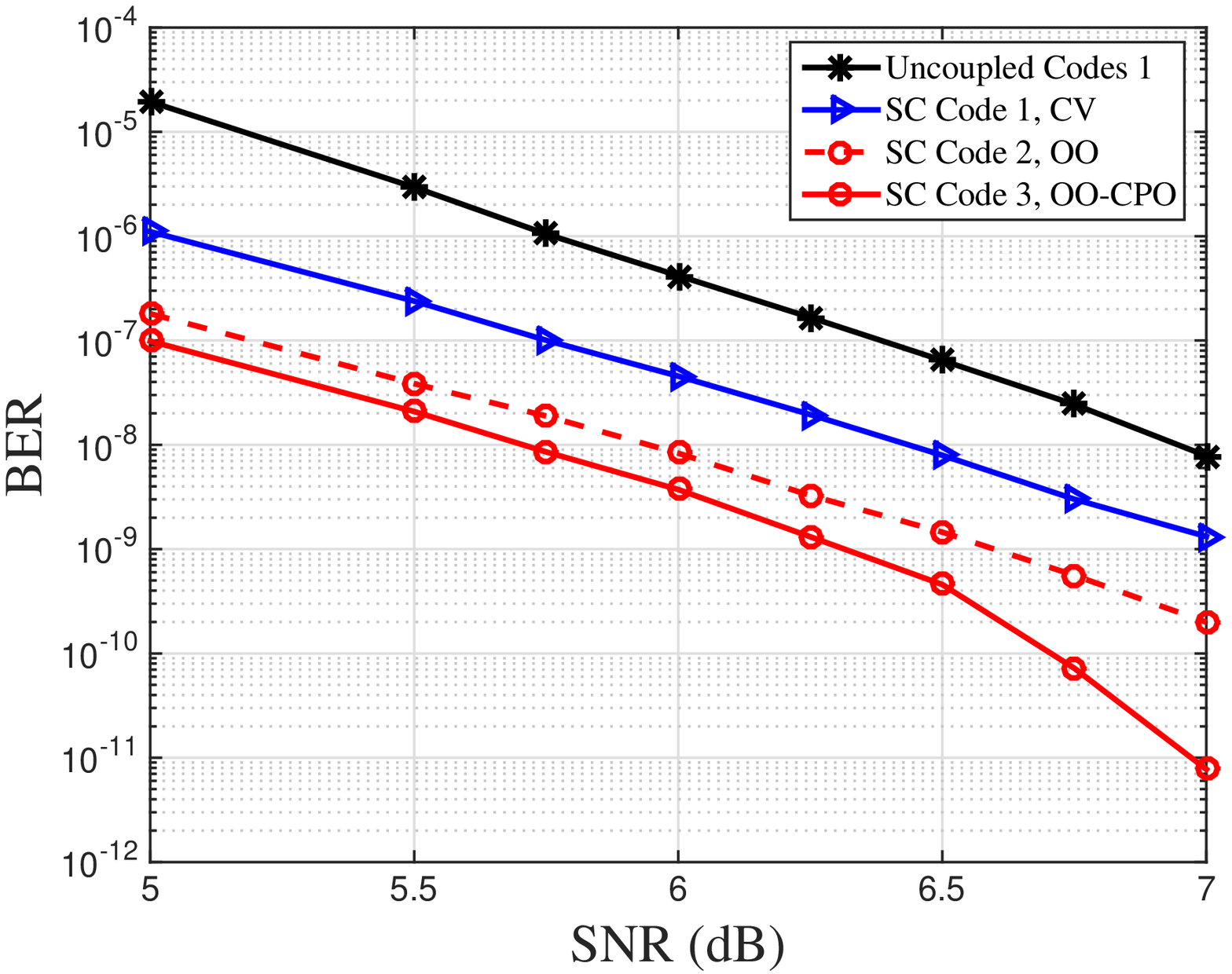}&\includegraphics[width=0.4\textwidth]{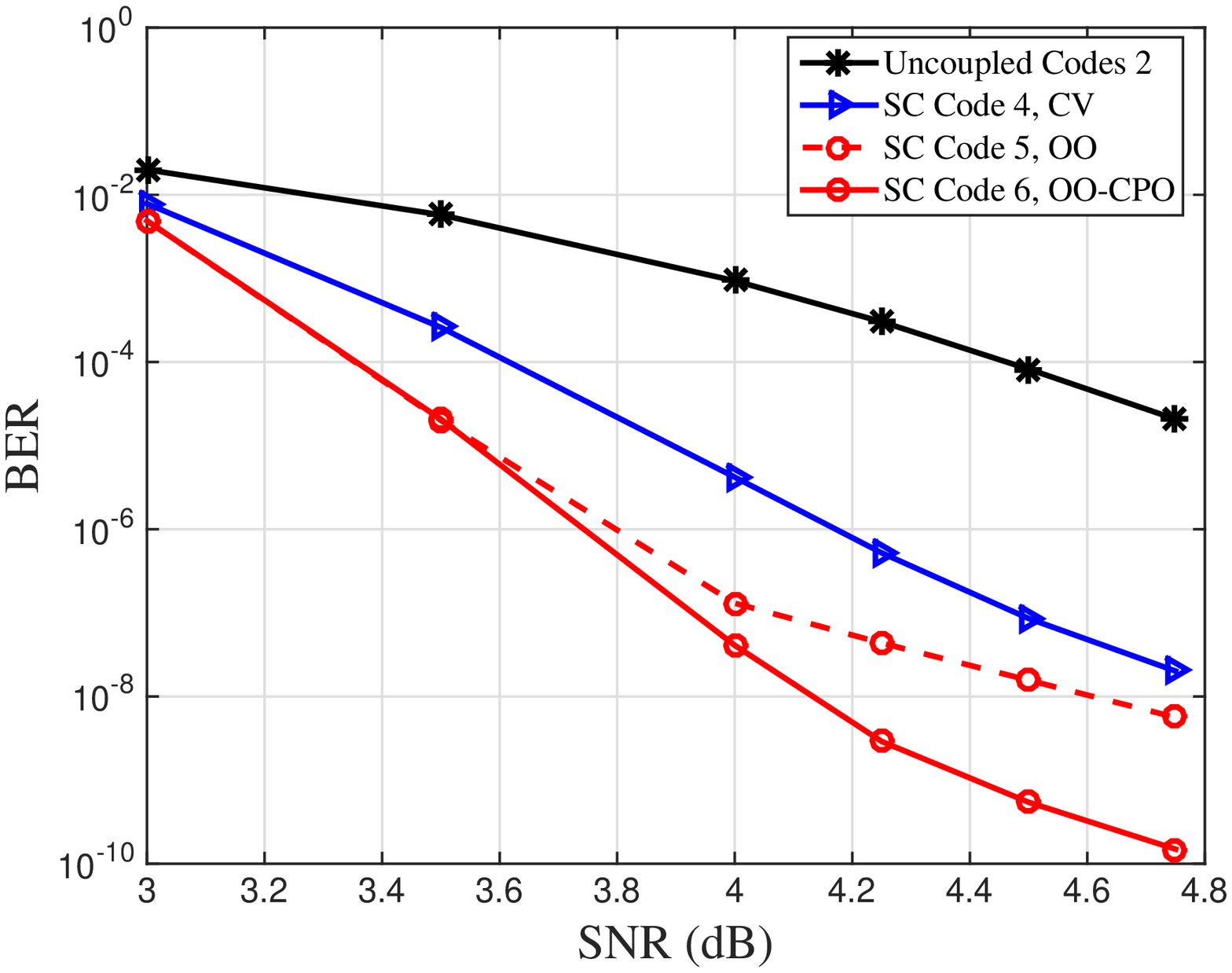}\vspace{-0.5cm}\\
\vspace{-0.2cm}
(a)&(b)\vspace{-0.4cm}
\end{tabular}
\caption{BER Curves over AWGN channel for SC codes with length $8,670$ bits, memory $m=1$, and constructed with different methods: (a) $\gamma=3$; (b) $\gamma=4$.\vspace{-0.3cm}}
\end{figure}
\begin{figure}\vspace{-0.8cm}
\centering
\includegraphics[width=0.4\textwidth]{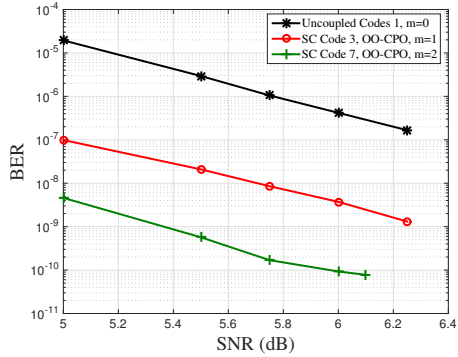}\vspace{-0.7cm}
\caption{BER Curves over AWGN channel for SC codes with length $8,670$ bits, $\gamma=3$, and different memories.\vspace{-1.1cm}}
\end{figure}
Fig.~9 shows the BER performance in the error floor area for SC codes with parameters $\kappa=p=17$, $\gamma=3$, $L=30$, constructed by the OO-CPO framework, and for different memories. Here, we consider the uncoupled case as an SC code with $m=0$ (no coupling). As we see, increasing the memory notably improves the error floor performance of SC codes.\vspace{-0.6cm}
\section{\vspace{-0.3cm}Conclusion}
In this paper, we presented a new combinatorial approach for the finite-length analysis and design of circulant-based SC codes.  We exploited the structure of SC codes to present a new methodology for the enumeration of combinatorial objects that can be applied to SC codes constructed by a wide variety of partitioning schemes and memories. Next, we introduced a novel partitioning scheme that operates on the protograph of an SC code to minimize the number of detrimental objects. Then, we proposed a heuristic approach for circulant power optimization that further reduces the population of these problematic objects in the final  graph. The proposed OO-CPO approach is an effective tool to construct SC codes that have a notably better error floor performance over other existing construction techniques. A promising research direction is to investigate the SC codes constructed by our new approach in modern dense storage applications.\vspace{-0.5cm}
\section{Appendices}
\vspace{-0.4cm}
\subsection{Proof of Lemma 3}
\vspace{-0.1cm}
\begin{IEEEproof}
From Definition~\ref{def_ov_par}, $t_{\{i_1,\cdots,i_{d_1},j_1,\cdots,j_{d_2}\}}$ is the number of overlaps (column indices in which all the rows $\{i_1,\cdots,i_{d_1},j_1,\cdots,j_{d_2}\}$ in $\mathbf{\Pi}_1^{1,p}$ have $1$'s simultaneously).
\begin{itemize}
\item $\mathcal{I}\neq \varnothing$:

To have an overlap at position (column index) $y\in\{1,\cdots\kappa\}$, 1) the rows in $\mathcal{I}$ of $\mathbf{\Pi}_1^{1,p}$ must have $1$'s at position $y$, 2) all rows in the first $m$ component matrices of $\mathbf{\Pi}_1^{1,p}$ corresponding to the rows in $\mathcal{J}$ must have $0$'s at position $y$. In other words, the rows $\{x_1\gamma+\overline{j_1},\cdots,x_{d_2}\gamma+\overline{j_{d_2}}\}$ of $\mathbf{\Pi}_1^{1,p}$ must have $0$'s at position $y$, where $[x_1,\cdots,x_{d_2}]\in\{0,\cdots,m-1\}^{d_2}$. Aided by the principle of inclusion-exclusion:
\begin{equation*}
t_{\mathcal{I}\cup\mathcal{J}}=t_\mathcal{I}+
\sum_{\alpha=1}^{d_2}{(-1)^\alpha}
\sum_{\substack{\{j_1',\cdots, j_\alpha'\}\subset\mathcal{J}\\ [ x_1,\cdots,x_\alpha ] \in\{0,\cdots,m-1\}^\alpha}}
t_{\mathcal{I}\cup\{x_1\gamma+\overline{j_1'},\cdots,x_{\alpha}\gamma+\overline{j_\alpha'}\}}.
\end{equation*}
\item $\mathcal{I}= \varnothing$:

To have an overlap at position (column index) $y\in\{1,\cdots\kappa\}$, all rows in the first $m$ component matrices of $\mathbf{\Pi}_1^{1,p}$ corresponding to the rows in $\mathcal{J}$ must have $0$'s at position $y$. In other words, the rows $\{x_1\gamma+\overline{j_1},\cdots,x_{d_2}\gamma+\overline{j_{d_2}}\}$ of $\mathbf{\Pi}_1^{1,p}$ must have $0$'s at position $y$, where $[x_1,\cdots,x_{d_2}]\in\{0,\cdots,m-1\}^{d_2}$. Aided by the principle of inclusion-exclusion:
\begin{equation*}
t_{\mathcal{J}}=\kappa+
\sum_{\alpha=1}^{d_2}{(-1)^\alpha}
\sum_{\substack{\{j_1',\cdots, j_\alpha'\}\subset\mathcal{J}\\ [ x_1,\cdots,x_\alpha ] \in\{0,\cdots,m-1\}^\alpha}}
t_{\{x_1\gamma+\overline{j_1'},\cdots,x_{\alpha}\gamma+\overline{j_\alpha'}\}}.
\end{equation*}
\end{itemize}\vspace{-0.4cm}
\end{IEEEproof}\vspace{-1.0cm}
\subsection{Proof of Lemma 5}
\vspace{-0.1cm}
\begin{IEEEproof}
A cycle-$6$ is formed of three distinct overlaps as illustrated in Fig.~5. To avoid over-counting, we must consider degree-$3$ overlaps in order to guarantee that the overlaps of a cycle-$6$ that belong to one replica have distinct indices.
\begin{enumerate}
\item 
Because of the structure of SC codes, an overlap between rows $((r-1)\gamma+i_1,(r-1)\gamma+i_2)$ in $\mathbf{R}_r$ corresponds to an overlap between rows $(i_1,i_2)$ in $\mathbf{\Pi}_1^{1,p}$. Similarly, an overlap between rows $((r-1)\gamma+i_1,(r-1)\gamma+i_3)$ in $\mathbf{R}_r$ corresponds to an overlap between rows $(i_1,i_3)$ in $\mathbf{\Pi}_1^{1,p}$, and an overlap between rows $((r-1)\gamma+i_2,(r-1)\gamma+i_3)$ in $\mathbf{R}_r$ corresponds to an overlap between rows $(i_2,i_3)$ in $\mathbf{\Pi}_1^{1,p}$. If $t_{\{i_1,i_2,i_3\}}=0$, the number of ways we can pick the overlaps is $t_{\{i_1,i_2\}}t_{\{i_1,i_3\}}t_{\{i_2,i_3\}}$. Since we must consider the degree-$3$ overlap among the rows $i_1$, $i_2$, and $i_3$ of $\mathbf{\Pi}_1^{1,p}$, we partition the enumeration into the following four cases:
\begin{itemize}
\item The overlap between rows $(i_1,i_2)$ is chosen from $t_{\{i_1,i_2,i_3\}}$ overlaps among the three rows, and the overlap between rows $(i_1,i_3)$ is chosen from other $(t_{\{i_1,i_2,i_3\}}-1)$ overlaps among the three rows (if possible; that is why we use $[t_{\{i_1,i_2,i_3\}}-1]^+$ ).
\item The overlap between rows $(i_1,i_2)$ is chosen from $t_{\{i_1,i_2,i_3\}}$ overlaps among the three rows, and the overlap between rows $(i_1,i_3)$ is chosen from $(t_{\{i_1,i_3\}}-t_{\{i_1,i_2,i_3\}})$ overlaps that are exclusively between these two rows.
\item The overlap between rows $(i_1,i_2)$ is chosen from $(t_{\{i_1,i_2\}}-t_{\{i_1,i_2,i_3\}})$ overlaps that are exclusively between these two rows, and the overlap between rows $(i_1,i_3)$ is chosen from $t_{\{i_1,i_2,i_3\}}$ overlaps among the three rows.
\item The overlap between rows $(i_1,i_2)$ is chosen from $(t_{\{i_1,i_2\}}-t_{\{i_1,i_2,i_3\}})$ overlaps that are exclusively between these two rows, and the overlap between rows $(i_1,i_3)$ is chosen from $(t_{\{i_1,i_3\}}-t_{\{i_1,i_2,i_3\}})$ overlaps that are exclusively between these two rows.
\end{itemize}\vspace{-0.08cm}
These four cases correspond to the four terms in (\ref{equ_A}).
\item Because of the structure of SC codes, an overlap between rows $((r-1)\gamma+i_1,(r-1)\gamma+i_2)$ in $\mathbf{R}_r$, an overlap between rows $((r-1)\gamma+i_1,(r-1)\gamma+i_3)$ in $\mathbf{R}_r$, and an overlap between rows $((r-1)\gamma+i_2,(r-1)\gamma+i_3)$ in $\mathbf{R}_q$ correspond to overlaps between pairs of rows $(i_1,i_2)$, $(i_1,i_3)$, and $(i_2+(r-q)\gamma,i_3+(r-q)\gamma)$ in $\mathbf{\Pi}_1^{1,p}$, respectively. The overlaps between rows $(i_1,i_2)$ and between rows $(i_1,i_3)$ belong to the same replica and must have distinct indices. The third overlap belongs to another replica and automatically has a distinct index. The number of options for the third overlap is $t_{i_2+(r-q)\gamma,i_3+(r-q)\gamma}$. We partition the enumeration into the following two cases:
\begin{itemize}
\item The overlap between rows $(i_1,i_2)$ is chosen from $t_{\{i_1,i_2,i_3\}}$ overlaps among the three rows, and the overlap between rows $(i_1,i_3)$ is chosen from the $[t_{\{i_1,i_3\}}-1]^+$ remaining options.
\item The overlap between rows $(i_1,i_2)$ is chosen from $(t_{\{i_1,i_2\}}-t_{\{i_1,i_2,i_3\}})$ overlaps that are exclusively between these two rows, and the overlap between rows $(i_1,i_3)$ is chosen from the $t_{\{i_1,i_3\}}$ options.
\end{itemize}
These two cases correspond to the two terms in (\ref{equ_B}).
\item Because of the structure of SC codes, an overlap between rows $((r-1)\gamma+i_1,(r-1)\gamma+i_2)$ in $\mathbf{R}_r$, $((r-1)\gamma+i_1,(r-1)\gamma+i_3)$ in $\mathbf{R}_q$, and $((r-1)\gamma+i_2,(r-1)\gamma+i_3)$ in $\mathbf{R}_s$ correspond to overlaps between pairs of rows $(i_1,i_2)$, $(i_1+(r-q)\gamma,i_3+(r-q)\gamma)$, and $(i_2+(r-s)\gamma,i_3+(r-s)\gamma)$ in $\mathbf{\Pi}_1^{1,p}$, respectively. These overlaps belong to different replicas and thus have distinct indices. Consequently, the number of cycles-$6$ in this case is the number of ways that we can choose these three overlaps, which is given in (\ref{equ_C}).\vspace{-0.4cm}
\end{enumerate}
\end{IEEEproof}\vspace{-0.8cm}
\subsection{\vspace{-0.1cm}Proof of Theorem 2}
\vspace{-0.1cm}
\begin{IEEEproof}
For a cycle-$6$, all the three VNs are adjacent (connected to each other via one distinct CN). As a result, for an SC code with memory $m$, a cycle-$6$ spans at most $\chi=m+1$ consecutive replicas, and (\ref{equ_enumeration_proto}) directly follows from (\ref{equ_enumeration}).
\begin{enumerate}
\item For $k=1$, we look for the number of cycles-$6$ having all the three VNs in replica $\mathbf{R}_1$ of $\mathbf{H}_{SC}^p$. Based on Lemma~\ref{lemma_A_B_C}, the number of cycles-$6$ with CNs $i_1$, $i_2$, and $i_3$ and VNs in $\mathbf{R}_1$ is $\mathcal{A}(t_{\{i_1,i_2,i_3\}},t_{\{i_1,i_2\}},t_{\{i_1,i_3\}},t_{\{i_2,i_3\}})$. Then, we just need to find all possible choices for the CNs $i_1$, $i_2$, and $i_3$. First, all the CNs must belong to the non-zero part of $\mathbf{R}_1$, i.e., $\{0,\cdots,(m+1)\gamma\}$. Second, the rows correspond to these CNs must have non-zero overlaps, i.e., $\overline{i_1}\neq\overline{i_2},\hspace{0.1cm}\overline{i_1}\neq\overline{i_3},\hspace{0.1cm}\overline{i_2}\neq\overline{i_3}$ (see Remark~\ref{remark_zero_ov_par_2}). Putting this together results in $F_1^1$ in (\ref{equ_enumeration_proto_Fk}).
\item For $k=2$, we look for the number of cycles-$6$ spanning the two replicas $\mathbf{R}_1$ and $\mathbf{R}_2$ of $\mathbf{H}_{SC}^p$, such that either two VNs are in $\mathbf{R}_1$ and one VN is in $\mathbf{R}_2$, or vice versa. Based on Lemma~\ref{lemma_A_B_C}, the number of cycles-$6$ with two VNs in $\mathbf{R}_1$, one VN in $\mathbf{R}_2$, and CNs $i_1$, $i_2$, and $i_3$ is $\mathcal{B}(t_{\{i_1,i_2,i_3\}},t_{\{i_1,i_2\}},t_{\{i_1,i_3\}},t_{\{i_2-\gamma,i_3-\gamma\}})$. Then, we just need to find all possible choices for CNs $i_1$, $i_2$, and $i_3$. Two CNs $i_2$ and $i_3$ are connected to VNs in replicas $\mathbf{R}_1$ and $\mathbf{R}_2$, thus they must belong to $\{\gamma,\cdots,(m+1)\gamma\}$, and CN $i_1$ is connected to VNs in $\mathbf{R}_1$, thus it must belong to $\{0,\cdots,(m+1)\gamma\}$. Putting this together results in the first summation $F_1^2$ in (\ref{equ_enumeration_proto_Fk}). The second summation, which is for the case when one VN belongs to $\mathbf{R}_1$ and two VNs belong to $\mathbf{R}_2$, can be found similarly.
\item For $3\leq k \leq (m+1)$, we look for the number of cycles-$6$ spanning $k$ replicas of $\mathbf{H}_{SC}^p$ starting from $\mathbf{R}_1$. Then, the first VN belongs to $\mathbf{R}_1$, the last VN belongs to $\mathbf{R}_k$, and the middle VN belongs to $\mathbf{R}_q$, ($1\leq q \leq k$). The first two summations in the expression of $F_1^k$ in (\ref{equ_enumeration_proto_Fk}) correspond to the cases $q=1$ and $q=k$ (the proof is similar to the previous case). For the case of $2\leq q \leq (k-1)$ and based on Lemma~\ref{lemma_A_B_C}, the number of cycles-$6$ with one VN in $\mathbf{R}_1$, one VN in $\mathbf{R}_q$, one VN in $\mathbf{R}_k$, and CNs $i_1$, $i_2$, and $i_3$ is $\mathcal{C}(t_{\{i_1,i_2\}},t_{\{i_1+(1-q)\gamma,i_3+(1-q)\gamma\}},t_{\{i_2+(1-k)\gamma,i_3+(1-k)\gamma\}})$. The CN $i_1$ is connected to VNs in $\mathbf{R}_1$ and $\mathbf{R}_q$, therefore it must belong to $\{(q-1)\gamma,\cdots,(m+1)\gamma\}$. The CN $i_2$ is connected to VNs in $\mathbf{R}_1$ and $\mathbf{R}_k$, therefore it must belong to $\{(k-1)\gamma,\cdots,(m+1)\gamma\}$. The CN $i_3$ connected to VNs in $\mathbf{R}_q$ and $\mathbf{R}_k$, therefore it must belong to $\{(k-1)\gamma,\cdots,(m+q)\gamma\}$. Putting this together results in the third summation of $F_1^k$ in (\ref{equ_enumeration_proto_Fk}).
\end{enumerate}\vspace{-0.5cm}
\end{IEEEproof}
\vspace{-0.6cm}
\bibliographystyle{IEEEtran}
\bibliography{IEEEabrv,references}

\end{document}